\documentclass[letter]{IEEEtran}
\usepackage{amsmath, amssymb, graphicx, color, cite, url}
\usepackage[latin1]{inputenc} \usepackage[T1]{fontenc}

\newcommand{\trp}[1]{{{\vphantom{#1}}^t#1}}

\newcommand{\bfi}{\bfseries\itshape}
\newcommand{\rem}[1]{}

\begin{document}
\topmargin 1cm
\textheight 21cm

\title{Soliton Dynamics in Computational Anatomy}
\author{ Darryl D. Holm${}^{1,2}$, J. Tilak Ratnanather${}^3$, Alain Trouv\'e${}^{5,6}$
and Laurent Younes${}^{3,4}$ \\\small
${}^{1}$Theoretical Division,
Los Alamos National Laboratory, Los Alamos, NM 87545, USA
\\ {\small dholm@lanl.gov}
\\${}^{2}$Mathematics Department,
Imperial College London, SW7 2AZ, UK
\\ {\small d.holm@imperial.ac.uk}
\\${}^3$ Center for Imaging Science, The Johns Hopkins University,
3400 N. Charles St., Baltimore MD 21218
\\{\small tilak@cis.jhu.edu, younes@cis.jhu.edu}
\\${}^4$ Department of Applied Mathematics and Statistics, the Johns Hopkins University
\\${}^5$CMLA (CNRS, UMR 8536), Ecole Normale Sup\'erieure de Cachan, 61,
\\Avenue du Pr\'esident Wilson, F-94 235 Cachan CEDEX, France
\\ {\small trouve@cmla.ens-cachan.fr}
\\${}^6$LAGA (CNRS, UMR 7539), Universit\'e Paris 13, Avenue J.-B. Clément, F-93430 Villetaneuse, France
}
\maketitle

\begin{abstract} \noindent
Computational Anatomy (CA) has introduced the idea of anatomical
structures being transformed by geodesic deformations on groups of
diffeomorphisms. Among these geometric structures, landmarks and
image outlines in CA are shown to be  singular solutions of a
partial differential equation that is called the geodesic {\em
EPDiff} equation. A recently discovered momentum map for singular
solutions of EPDiff yields their canonical Hamiltonian
formulation, which in turn provides a complete parameterization of
the landmarks by their canonical positions and momenta. The
momentum map provides an {\em isomorphism} between landmarks (and
outlines) for images and singular soliton solutions of the EPDiff
equation. This isomorphism suggests a new dynamical paradigm for
CA, as well as new data representation.
\end{abstract}

%
%

\section{Introduction}
Computational Anatomy (CA) must measure and analyze a range of
variations in shape, or appearance, of highly deformable
structures. Following the pioneering work by Bookstein, Grenander
and Bajscy
\cite{bookstein-1991,grenander-book-1993,bajscy-etal-1983}, the
past several years have seen an explosion in the use of template
matching methods in computer vision and medical imaging
\cite{toga-1999,thompson-etal-2000,toga-thompson-2003,ashburner-etal-2003,DuGrMi1998,hallinan,miller-younes-2001,Tr1995,Tr1998,GrMi1998,miller-etal-2002,mumford-pattern,mumford-elastica,mumford-shape,jain-etal-1998,montagnat-etal-2001}.
These methods have enabled the systematic measurement and
comparison of anatomical shapes and structures in biomedical
imagery leading to better understanding of neurodevelopmental,
neuropsychiatric and neurological disorders in recent years
\cite{jgc-2004,posener-2003,tepest-2003,wang-2003,sowell-2003-lancet,thompson-2003,narr-2004-hippo-sz,sowell-2003,narr-2003-csf,narr-2004,narr-2003-sz,gogtay-2004,gee-2003,cannon-2003,ballmaier-2004,ballmaier-2004-cingulate}.
The mathematical theory of Grenander's deformable template models,
when applied to these problems, involves smooth invertible maps
(diffeomorphisms), as presented in this context in
\cite{Tr1995,Tr1998}, \cite{DuGrMi1998}, \cite{mumfordihp},
\cite{miller-younes-2001} and \cite{miller-etal-2002}. In
particular, the template matching approach involves Riemannian
metrics on the diffeomorphism group and employs their projections
onto specific landmark shapes, or image spaces.

On the other hand, the diffeomorphism group has also been the focus of
special attention in fluid mechanics. For example, Arnold \cite{Ar1966}
proved that ideal incompressible fluid flows correspond to geodesics on the
diffeomorphism group, with respect to the metric provided by the fluid's
kinetic energy. In this paper, we shall draw parallels between these two
endeavors, by showing how the Euler-Poincaré theory of ideal fluids can
be used to develop new perspectives in CA. In particular, we discover that
CA may be informed by the concept of canonical momentum for geodesic
flows describing the interaction dynamics for singular solitons in
shallow water called peakons
\cite{CaHo1993}.

\noindent
{\bfi Outline of the paper.} Section \ref{sec:comp.an}
describes the template matching variational problems of computational
anatomy, and introduces the fundamental EPDiff evolution equation,
which describes the evolution of the momentum of an anatomy, (a
collection of landmarks, for example).  The singular solutions for the
EPDiff equation (\ref{EP-eqn}) with diffeomorphism group $G$ are explained
in section \ref{EPDiff-sing-soln-sec}. They are, in particular, related to the
landmark matching problem in computer vision. The consequences of
EPDiff for computational anatomy are described in section \ref{sec:cons}.
Conclusions are summarized in section \ref{conc-sec}.

%
\section{Variational Formulation of Template Matching Problem}
\label{sec:comp.an}
\subsection{Geometrical large deformation setting.}
Grenander \cite{GrMi1994} pioneered the introduction of group
actions in image analysis, through the notion of {\em deformable
templates}. Roughly speaking, a deformable template is an ``object, or
exemplar'' $I_0$ on which a group ${\mathcal G}$ acts and thereby generates,
through the orbit ${\mathcal I} = {\mathcal G}I_0$, a family of new objects.
This ``template matching'' approach has proven to be versatile and useful
in different settings (image matching, landmark matching, surface matching
and more recently in several extensions of metamorphoses \cite{TrYo2004})
The approach focuses its modeling effort on properties of the families of
shapes generated by the action of the group ${\mathcal G}$ on the deformable
templates. Right or left invariant geodesic distances on the group
${\mathcal G}$ are the {\em natural} extension to large deformation of the
quadratic cost, or effort function, defined for  small
linearized perturbations of the identity element.

\subsection{Case of non-rigid template matching.}
\label{par:var} The optimal solution to a non-rigid template matching
problem is the shortest, or least expensive, path of continuous
deformation of one geometric object (template ) into another one (target).
For this purpose, we have introduced a time-indexed deformation process,
starting at time $t=0$ with the template (denoted $I_0$), and reaching the
target at time $t=1$. At a given time $t$ during this process, the current
object
$I_t$ is assumed to be the image of the template, $I_0$, through
the (left) action of a diffeomorphism $\phi_t$: $I_t = \varphi_t\cdot I_0$.
The attribution of a cost to this process is then based on functionals
defined on the group of diffeomorphisms, following Grenander's
principles.

A simple and natural way to assign a cost to a diffeomorphic process
indexed by time is based on the following: to measure the
variation $\varphi_{t+dt} - \varphi_t$, express this difference
as a small vector of displacement, $dt\, {\mathbf{u}}_t$, composed
with $\varphi_t$,
$$
\varphi_{t+dt} - \varphi_t = dt\, {\mathbf{u}}_t \circ \varphi_t
$$
The cost of this small variation is then expressed as a function of
${\mathbf{u}}_t$ only, yielding the final expression,
$$
\text{Cost}(t\mapsto\varphi_t) = \int_0^1 \ell({\mathbf{u}}_t) \,dt
$$
with
\begin{equation}
\label{eq:1}
\frac{d\phi_t}{dt} = {\mathbf{u}}_t \circ \varphi_t
\end{equation}

In the following, the function ${\mathbf{u}}_t\mapsto
\ell({\mathbf{u}}_t)$ is defined as a squared functional norm on the
infinite dimensional space of velocity vectors. This process is a standard
construction in the Riemannian geometry of Lie groups, in which the
considered group is equipped with a right invariant Riemannian metric.
Here, the vector space of right invariant instantaneous velocities,
${\mathbf{u}}_t=(d\phi_t/dt)\circ\varphi_t^{-1}$, forms the tangent space at
the identity of the considered group, and may be regarded as its {\em Lie
algebra}. This vector space will be denoted $\mathfrak{g}$ in the
following (as a formal analogy Lie algebra notation). In
this context, the cost of a time-dependent deformation process, thus
defined by
\begin{equation}
\label{eq:ener}
\text{Cost}(t\mapsto\varphi_t) = \int_0^1
|{\mathbf{u}}_t|_{\mathfrak{g}}^2\,dt
\,,
\end{equation}
is the {\em geodesic action} of the process for this Riemannian metric,
and most problems in CA can be formulated as {\em
  finding the deformation path with minimal action, under the constraint
  that it carries the template to the target}. We will illustrate this
below with an important example of a such problem, in which the
geometric objects are collections of points in space (landmarks).
Before doing this, we summarize in more rigorous terms the process
described above, providing at the same time the notation to be
used in the rest of the paper. However, most of the remaining of
the paper can be understood by referring to the summary paragraph
at the end of section \ref{sec:rig}.

\subsection{Rigourous construction.}
\label{sec:rig}
Fix an open, bounded subset
$\Omega \subset {\mathbb R}^d$.  Following \cite{Tr1995}, the construction
 is based on the design of the ``Lie algebra'' $\mathfrak{g}$,
which is in turn used to generate the group elements. (This is the
converse of the usual consideration of finite dimensional Lie
groups.) The following construction of $\mathfrak{g}$ will be
assumed. Denote by $H$ the set of square integrable vector fields
on $\Omega$ with the usual $L^2$ metric $(\,\cdot,\cdot\,):\,H\times
H\to\mathbb{R}$. Consider a symmetric and coercive operator
$L:{\mathbf{u}}\mapsto L{\mathbf{u}} \in H^*=H$ whose domain $D(L)$
contains all smooth ($C^\infty$) vector fields with compact support in
$\Omega$. This operator induces an inner product on
$D(L)$ by $\langle {\mathbf{u}},{\mathbf w}\rangle_{\mathfrak g}=
(L{\mathbf{u}},{\mathbf w})$. This pre-Hilbert space can be
completed to form a Hilbert space (\cite{Ze1995}), thereby defining
${\mathfrak{g}}$ which is continuously embedded in $H$
(Freidrich's extension). Suppose, in addition, that ${\mathfrak{g}}$ can
be embedded into $C^p(\Omega)$, the set of $p$ times continuously
differentiable vector fields on $\Omega$, with
$p\geq 1$. (We call this the {\em $p$-admissibility condition}.) Then the
following can be shown (\cite{Tr1995,DuGrMi1998}):
{\em If ${\mathbf{u}}_t$ is a time-dependent family of elements
of ${\mathfrak{g}}$ such that
$\int_0^1 \|{\mathbf{u}}_t\|^2_{\mathfrak g} dt < \infty$, that is,
${\mathbf{u}}_t\in L^2([0,1], {\mathfrak{g}})$,
then the flow $\frac{\partial \varphi}{\partial t} = {\mathbf u}_t
\circ \varphi_t$
with initial conditions $\varphi_0(x)= x, \ x\in\Omega$, can be
integrated over $[0,1]$, and $\varphi_1$ is a diffeomorphism of
$\Omega$, which is denoted $\varphi^{\mathbf{u}}_1$.} The image of
$L^2([0,1], {\mathfrak g})$ by ${\mathbf u} \mapsto
\varphi^{\mathbf{u}}_1$ forms our group of diffeomorphisms, ${\mathcal G}$,
which is therefore {\em  specified by the operator $L$}. In this setting,
equation \eqref{eq:1} has solutions over any  finite interval,
and the infinum
should be taken with respect to
$\varphi=\varphi^{\bf u}\circ\varphi_0$ for ${\bf u}\in
L^2([0,1],{\mathfrak g})$ such that
$\varphi_1=\varphi^{\bf
  u}_1\circ\varphi_0$. (Because of right-invariance, it can
furthermore be assumed that $\phi_0 = \text{id}$.) Moreover, as proved
in \cite{Tr1995, DuGrMi1998}, the existence
of a minimizer (a geodesic path) is guaranteed.

For the inner product $\langle {\mathbf{u}},{\mathbf w}\rangle_{\mathfrak g}
 =
(L{\mathbf{u}},{\mathbf w})$, the operator $L$ is a
{\em duality map}. In a deformation process $t\mapsto \varphi_t$ such that
$$
\frac{d\phi_t}{dt} = {\mathbf u}_t \circ \varphi_t
$$
${\mathbf u}_t$ is called the (Eulerian) {\em velocity}, belonging to $\mathfrak{g}$,
and $L{\mathbf u}_t$ is called the {\em momentum}, also denoted
${\mathbf{m}}_t$. Note
that, because $L$ will typically be a differential operator, the
momentum $L{\mathbf u}$ can (and will) be singular, e.g. a measure, or a
generalized function. A simple example of such a phenomenon occurs in the
landmark matching problem described in the next subsection.

This framework for CA is reminiscent of the least action
principle for continuum motion of fluids with Lagrangian $\ell({\bf
u})=\frac{1}{2}\|{\bf u}\|_{\mathfrak g}^2$.
Note that in the template matching framework, $\ell$ has the specific
interpretation of an effort functional for small deformations that
should be {\em designed} according to a given application and not follow
any existing physical model. However, this similarity
with ideal fluid dynamics sets the stage for ``technology transfer''
between computational image science and fluid dynamics -- e.g.,
Hamiltonian description, momentum evolution, classification of
equilibria, nonlinear stability, PDE analysis, etc. In particular,
the least action interpretation of \eqref{eq:ener}, is central in the
Arnold theory of hydrodynamics \cite{Ar1966} and the derivation of
the geodesic evolution equations falls into the
Euler-Poincaré (EP) theory, which produces the {\bfi EP motion
equation} \cite{HoMaRa1998, mumfordihp},
\begin{equation}\label{EP-eqn}
\Big(\frac{\partial}{\partial t}
+ {\mathbf{u}}\cdot\nabla\Big)
{\mathbf{m}}
+\,
\nabla \trp{\mathbf{u}}\cdot{\mathbf{m}}
+
{\mathbf{m}}({\rm div\,}{\mathbf{u}})
=
0
\,,
\end{equation}
and ${\mathbf{u}}=G*{\mathbf{m}}$, where $G*$ denotes convolution
with the Green's kernel $G$ for the operator $L$. This is the
{\bfi EPDiff equation}, for ``Euler-Poincaré equation on the
diffeomorphisms''.

\subsubsection*{Summary} The important consequences of the previous
construction is that, by measuring the amount of fluid deformation
which is required to morph an object to another, this measure being
given by \eqref{eq:ener} where $u_t$ is the velocity of the fluid
deformation at time $t$ and $|{\mathbf{u}}_t|_{\mathfrak{g}}^2 =
(Lu_t, u_t)$, $L$ being a linear operator, the associated momentum,
$m_t = Lu_t$ satisfies the Euler Poincaré equation
\eqref{EP-eqn}. This equation is important, because it allows to
reconstruct the complete evolution of the momentum (and hence of the
fluid motion) from the initial conditions. This property is exploited
for the analysis of landmark data in \cite{mtvy04}.

\subsection{Landmark matching and measure-based momentum}
The landmark matching problem is an interesting illustration of
the singularity of the momentum which naturally emerges in the
computation of geodesics. Given two collections of points $X_1,
\ldots, X_K$ and $Y_1, \ldots, Y_K$ in $\Omega$, the problem
consists in finding a time-dependent diffeomorphic process
$(t\mapsto \varphi_t)$ of minimal action (or cost, as given
by \eqref{eq:ener}) such that $\varphi_0 = \text{id}$ and
$\varphi_1(X_i) = Y_i$ for $i= 1, \ldots, K$. This problem was first
addressed in \cite{josmim00}, then studied in different forms
in \cite{CaYo01, glaunes03:sphere, beg-thesis, GlTrYo04}.  Its computational solution relies on the key observation that the
problem can be expressed uniquely in terms of optimizing the {\em landmark
trajectories}, $\mathbf Q_i(t) = \varphi_t(X_i)$, for an action, or cost,  given
by
\begin{equation}
\label{eq:land.en}
S
= \int_0^1 \ell(\mathbf Q, \mathbf Q^\prime) \,dt
= \frac{1}{2}\int_0^1 \trp{\mathbf Q^\prime}(t) A(\mathbf Q(t))^{-1}
\mathbf Q^\prime(t) \,dt
\,,
\end{equation}
with notation $\mathbf Q^\prime(t)=d\mathbf Q/dt$.
This action, or cost, $S=\int\ell({\bf u})\,dt$ is the time
integrated Lagrangian in the least action principle, $\delta S =0$.
Its end point conditions are $\mathbf Q_i(0) = X_i$ and $\mathbf Q_i(1) = Y_i$,
where $A(\mathbf Q)$ is an $Kd\times Kd$ matrix ($d$ is the dimension of the
underlying space) which may be constructed as follows. Let $G$ be the
Green's kernel associated to the operator $L$, formally defined by
$$
\mathbf v(x) = \int_\Omega G(x,y) (L\mathbf v)(y) dy
\,.$$
Let $I_d$ be the $d$-dimensional identity matrix. Then $A(\mathbf Q)$ is a
block matrix $(A_{ij}(\mathbf Q), i, j=1, \ldots, K)$ with
$A_{ij}(\mathbf Q) = G(\mathbf Q_i,
\mathbf Q_j) I_d$.

Denote $\mathbf P(t) = A(\mathbf Q(t))^{-1}\mathbf
Q^\prime(t)=\partial\ell/\partial \mathbf Q^\prime$.
Then, the optimal diffeomorphism $t\mapsto \varphi_t$ is given by equation
\eqref{eq:1} with
$$
\mathbf u_t(x) = \sum_{i=1}^K G(x, \mathbf Q_i(t))\mathbf P_i(t)\,.
$$
The corresponding momentum $m_t$ is given by the point measure
$$
\mathbf m_t(y) = \sum_{i=1}^K \mathbf P_i(t) \delta(y - \mathbf Q_i(t))\,.
$$
A straightforward extension of this model occurs when the landmarks are
organized along continuous curves in which the indices $i, j$ are
replaced by curve parameters (say defined over $[0,1]$), and
$$
\mathbf m_t(y) = \int_0^1 \mathbf P(t, s) \delta(y - \mathbf Q(t, s)) ds\,.
$$
One could also distribute the landmarks along
{\it several} continuous curves (the outlines of an image, say). This
representation of momentum would involve both integrations and sums.

As we will see, this measure-based momentum is also found in other
contexts very different from medical imaging. We also mention an
important variant of this matching problem. This variant is a
particular case of a general framework, in which the diffeomorphic
group action is extended to incorporate possible variation in the
template itself. The application of this theory of
``metamorphoses'' \cite{TrYo2004} to the particular case of
landmark matching simply comes with the addition of a new
parameter $\sigma^2 > 0$ and the replacement in the above formulas
of $A(\mathbf Q)$ by $A^\sigma(\mathbf Q) = A(\mathbf Q) +
\sigma^2 I_{Nd}$ ($I_{Nd}$ is the $Nd$ dimensional identity
matrix). This corresponds to geodesic interpolating splines
introduced in \cite{CaYo01}. Note that in this case, EPDiff is not
satisfied anymore, but may be modified to include a non-vanishing
right-hand term.


\section{EPDiff and its singular solutions}\label{EPDiff-sing-soln-sec}

\subsection{The EPDiff equation.}

The EPDiff equation is
important in fluid dynamics, because it encodes the fundamental
dynamical properties of energy, circulation and potential vorticity.
A first example comes with choosing the differential operator $L$
as $L{\mathbf u} = \mathbf u$, so that
$\mathbf m=\mathbf u$,  and taking incompressible vector fields,
so that $\text{div\,}{\mathbf{u}}=0$. In this setting,
equation \eqref{EP-eqn} provides the Euler equations for the
incompressible flow of an ideal fluid
\cite{Ar1966,ArKh1998}.

Another physically relevant form of the EPDiff equation is the
evolutionary integral-partial differential equation (\ref{EP-eqn}) with
\cite{HoMaRa1998,HoSt2003,HoMa2004}
\begin{equation}\label{Helmholtz-op}
{\mathbf{m}}
\equiv
L{\bf u}
=
{\mathbf{u}}
-
\alpha^2\Delta{\mathbf{u}}
\,,
\end{equation}
so that
$$|\mathbf u|_{\mathfrak g}^2 = \|{\mathbf{u}}\|_{H^1_\alpha}^2 = \int_\Omega (|\mathbf u|^2 + \alpha^2
|\nabla \mathbf u|^2) dx\,.$$
In this particular case of EPDiff, denoted as EPDiff(H1), one obtains the
velocity ${\mathbf{u}}$ from the momentum ${\mathbf{m}}$ by an inversion
of the elliptic Helmholtz operator $\big(1-\alpha^2\Delta)$, with length
scale $\alpha$ and Laplacian operator $\Delta$.
The EPDiff(H1) equation with momentum
definition (\ref{Helmholtz-op}) therefore describes {\it geodesic} motion on the
diffeomorphism group with respect to
$\|{\mathbf{u}}\|_{H_\alpha^1}^2$, the
$H_\alpha^1$ norm of the fluid velocity.
This velocity norm is recognized as being (twice) the ``kinetic
energy,'' when equation (\ref{EP-eqn}) with momentum definition
(\ref{Helmholtz-op}) is interpreted as a model fluid equation, as
in shallow water wave theory  \cite{HoMaRa1998,HoSt2003,HoSt2004}.
Although this operator does not appear in the class of operators
used in CA (because it does not satisfy the $p$ admissibility condition),
this model exhibits a number of features which are highly relevant also in
this case. In particular, singular momentum solutions emerge in the
initial value problem for this model which behave as isolated waves,
called solitons. In many ways, {\em landmark matching in CA can be seen
as generating a soliton dynamics between two sets of landmarks.}
\subsection{Singular momentum solutions of EPDiff.}
In the 2D plane, EPDiff, (\ref{EP-eqn}), has weak {\it
singular momentum solutions} that are expressed as
\cite{HoSt2003,HoMa2004}
\begin{equation}\label{EP-sing-mom}
{\mathbf{m}}({\mathbf{x}},t)
=
\sum_{a=1}^N\int_{s}
{\mathbf{P}}_a(t,s)\delta\big({\mathbf{x}}-{\mathbf{Q}}_a(t,s)\big)\,ds
\,,
\end{equation}
where $s$ is a {\bfi Lagrangian coordinate} defined along a set of
$N$ curves in the plane {\it moving with the flow} by the equations
${\mathbf{x}}={\mathbf{Q}}_a(t,s)$ and supported on the delta functions in
the EPDiff solution (\ref{EP-sing-mom}).
Thus, the singular momentum solutions of EPDiff
are vector valued curves supported on the delta functions in
(\ref{EP-sing-mom}) representing evolving ``wavefronts'' defined by the
{\bfi Lagrange-to-Euler map} (\ref{EP-sing-mom}) for their
momentum. These solutions have the exact same form as the landmark
solutions obtained in the previous section (with the straightforward
extension of matching $2N$ curves instead of 2).

Substituting the defining relation ${\mathbf{u}}\equiv
G*{\mathbf{m}}$ into the singular momentum solution (\ref{EP-sing-mom})
yields the velocity representation for the wavefronts, as another
superposition of integrals,
\begin{equation}\label{EP-sing-vel}
{\mathbf{u}}({\mathbf{x}},t)
=
\sum_{a=1}^N\int_{s}
{\mathbf{P}}_a(t,s)
G\big({\mathbf{x}},{\mathbf{Q}}_a(t,s)\big)\,ds
\,.
\end{equation}
In the example of EPDiff(H1), the Green's function $G$ for the
second order Helmholtz operator in  (\ref{Helmholtz-op}) relates
the velocity to the momentum. In this case, the velocity in
the singular solution (\ref{EP-sing-vel}) has a discontinuity in
its first derivative (its slope) across each curve parameterized
by $s$ moving with the flow. Being discontinuities in the gradient
of velocity that move along with the flow, these singular
solutions for the velocity are classified as  {\bfi contact
discontinuities} in fluid mechanics. These contact discontinuities
do not occur if the Green's kernel is sufficiently smooth (eg. a Gaussian
kernel), as is typically used in CA.

\subsection{Lagrangian representation of the singular solutions of EPDiff.}
\label{mom-map-subsec}

Substituting the singular momentum solution formula (\ref{EP-sing-mom})
for $s\in{\mathbb{R}}^1$ and its corresponding velocity
(\ref{EP-sing-vel}) into EPDiff (\ref{EP-eqn}), then integrating
against a smooth test function implies the following {\bfi Lagrangian wavefront equations}
\begin{eqnarray}
\frac{\partial }{\partial t}{\mathbf{Q}}_a (s,t)
&=&
\!\!\sum_{b=1}^{N} \int{\mathbf{P}}_b(s^{\prime},t)\,
G({\mathbf{Q}}_a(s,t),{\mathbf{Q}}_b(s^{\prime},t)\,\big)ds^{\prime}
\,,\nonumber\\
 \frac{\partial }{\partial t}{\mathbf{P}}_a (s,t)
&=&
-\,\!\!\sum_{b=1}^{N} \int
\trp{\mathbf{P}}_a(s,t)\,{\mathbf{P}}_b(s^{\prime},t)
\label{IntDiffEqn-Q}\\&&\hspace{1cm}
\, \frac{\partial }{\partial {\mathbf{Q}}_a(s,t)}
G\big({\mathbf{Q}}_a(s,t),{\mathbf{Q}}_b(s^{\prime},t)\big)\,ds^{\prime}
\,.
\nonumber
\end{eqnarray}
Thus, the momentum solution formula (\ref{EP-sing-mom})
yields a closed set of integral partial differential equations
given by  (\ref{IntDiffEqn-Q}) for the vector parameters
${\mathbf{Q}}_a(s,t)$ and ${\mathbf{P}}_a(s,t)$ with $a=1,2\dots N$.
The dynamics (\ref{IntDiffEqn-Q}) for these parameters is canonically
Hamiltonian and geodesic in phase space.

\subsection{Relation between contact solutions of EPDiff(H1) and solitons.}
As we have discussed, the weak solutions of EPDiff(H1) represent the third
of the three known types of fluid singularities: shocks, vortices and
contacts. The key feature of these contacts is that they carry momentum;
so the wavefront interactions they represent are {\it collisions}, in
which momentum is exchanged. This is very reminiscent of the soliton
paradigm in 1D. And, indeed, in 1D the singular
solutions (\ref{EP-sing-vel}) of EPDiff are true solitons that undergo
elastic collisions and are solvable by the inverse scattering transform
for an isospectral eigenvalue problem \cite{CaHo1993}. Besides describing
wavefronts, this interaction of contacts applies in a variety of fluid
situations ranging from solitons \cite{CaHo1993}, to turbulence
\cite{Chen-etal1998,FoHoTi2001}. The nonlocal elliptic solve
${\mathbf{u}}=G*{\mathbf{m}}$ relating the momentum density ${\mathbf{m}}$
to the velocity ${\mathbf{u}}$ in the EPDiff(H1) equation also appears in
the theory of fully nonlinear shallow water waves
\cite{CaHo1993,GrNa1976,HoSt2003,HoSt2004,SuGa1969}.

\begin{figure*}
\begin{center}
\includegraphics[width=12cm]{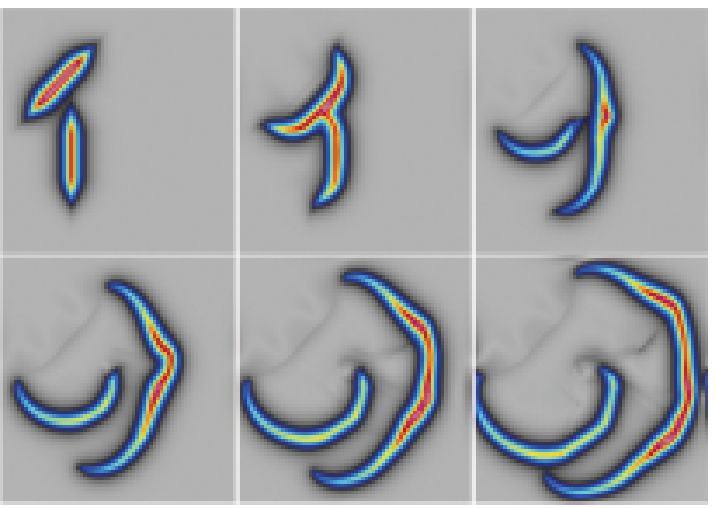}
\end{center}
\caption{
\label{reconnex-fig}
A single collision is shown to produce reconnection as the faster
wave front segment initially moving Southwest along the diagonal expands,
curves  and obliquely overtakes the slower one, which was initially moving
rightward (East). This reconnection illustrates one of the collision rules
for singular solutions of the two-dimensional EPDiff flow. See
\cite{HoSt2004} for a complete treatment.}

\end{figure*}

The physical concept of momentum exchange is well understood for
nonlinear collisions of shallow water waves, especially in 1D.
Momentum exchange for EPDiff in 1D is exhaustively studied in
\cite{FrHo2001}. The corresponding momentum exchange processes
(wavefront collisions) for EPDiff(H1) in 2D and 3D are studied in
\cite{HoSt2003,HoSt2004}. These wavefront collisions show an
interesting phenomenon. Namely, wavefront solutions of EPDiff(H1)
in 2D and 3D for which a faster wavefront obliquely overtakes a
slower one result in the faster wavefront accelerating the slower
one and {\it reconnecting} with it. Such wavefront reconnections
are observed in nature. For example, observations from the Space
Shuttle show internal wavefront reconnections occurring in the
South China Sea \cite{Liu-etal[1998]}. This wave front reconnection
phenomenon is illustrated in Figure \ref{reconnex-fig}. The key
mathematical feature responsible for wave front
reconnection is the nonlocal nonlinearity appearing in EPDiff.



\subsection{$H^1_\alpha$ norm versus smooth kernels}
As noted before, the Green's kernels for the operators
$L$ used in CA are typically smoother than the inverse
of the elliptic Helmholtz operator $(1-\alpha^2 \Delta)$ which
corresponds to the $H^1_\alpha$-model. A consequence of this is
that the (variational) matching problems are always well-posed in
CA, and their solutions are computationally feasible.

To illustrate the differences introduced by using smooth Green's kernels,
consider the case of a symmetric head-on collision of two particles (or landmarks)
in 1D. Under the $H^1_\alpha$ model, they will meet in finite time, then
bounce back after exchanging their momenta. This is impossible with a
smooth kernel, since the landmarks are carried by a diffeomorphic
motion, and therefore cannot meet if they started from different
positions.

\subsection{Metamorphoses}
\begin{figure*}
\begin{center}
\includegraphics[width=6cm]{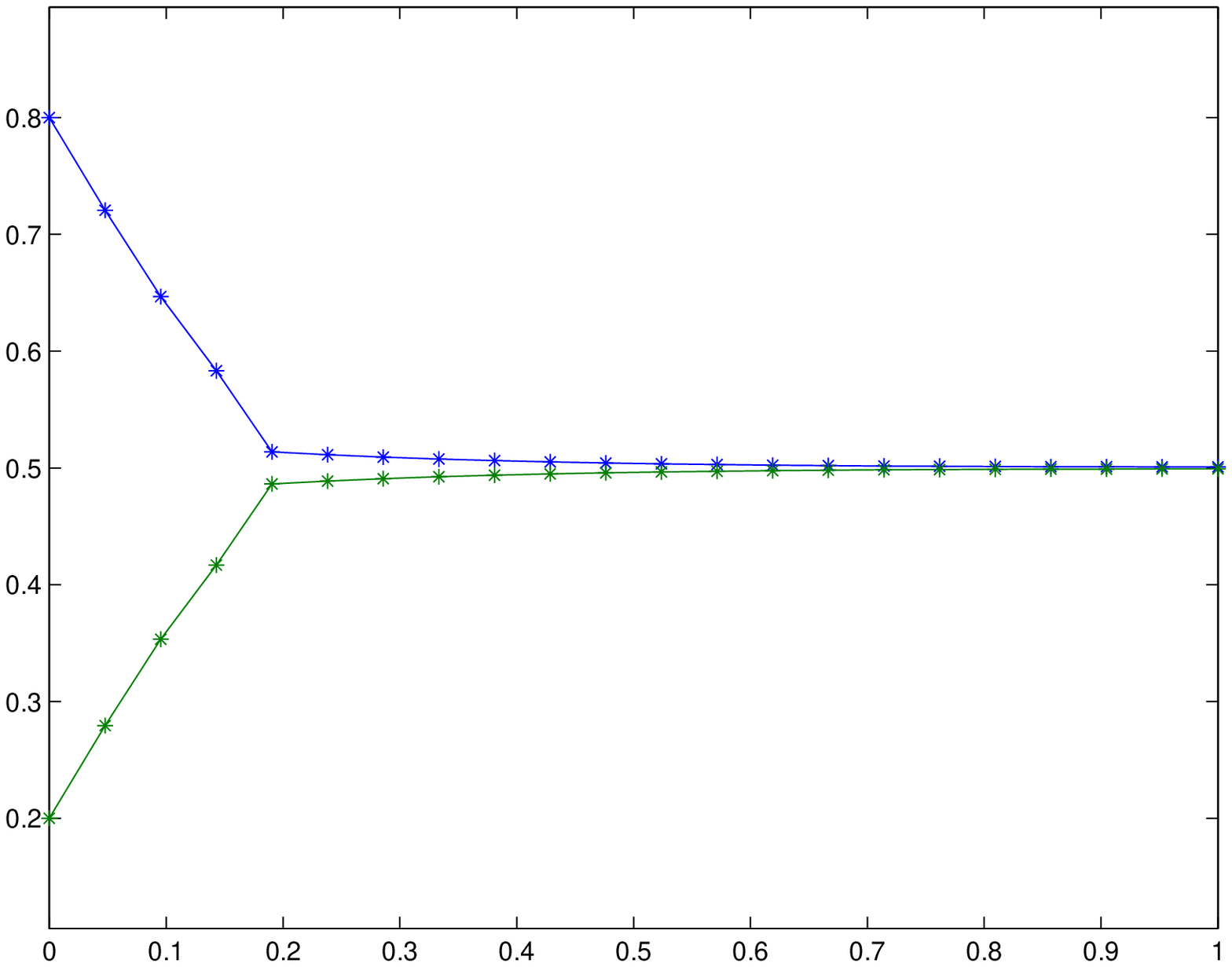}
\includegraphics[width=6cm]{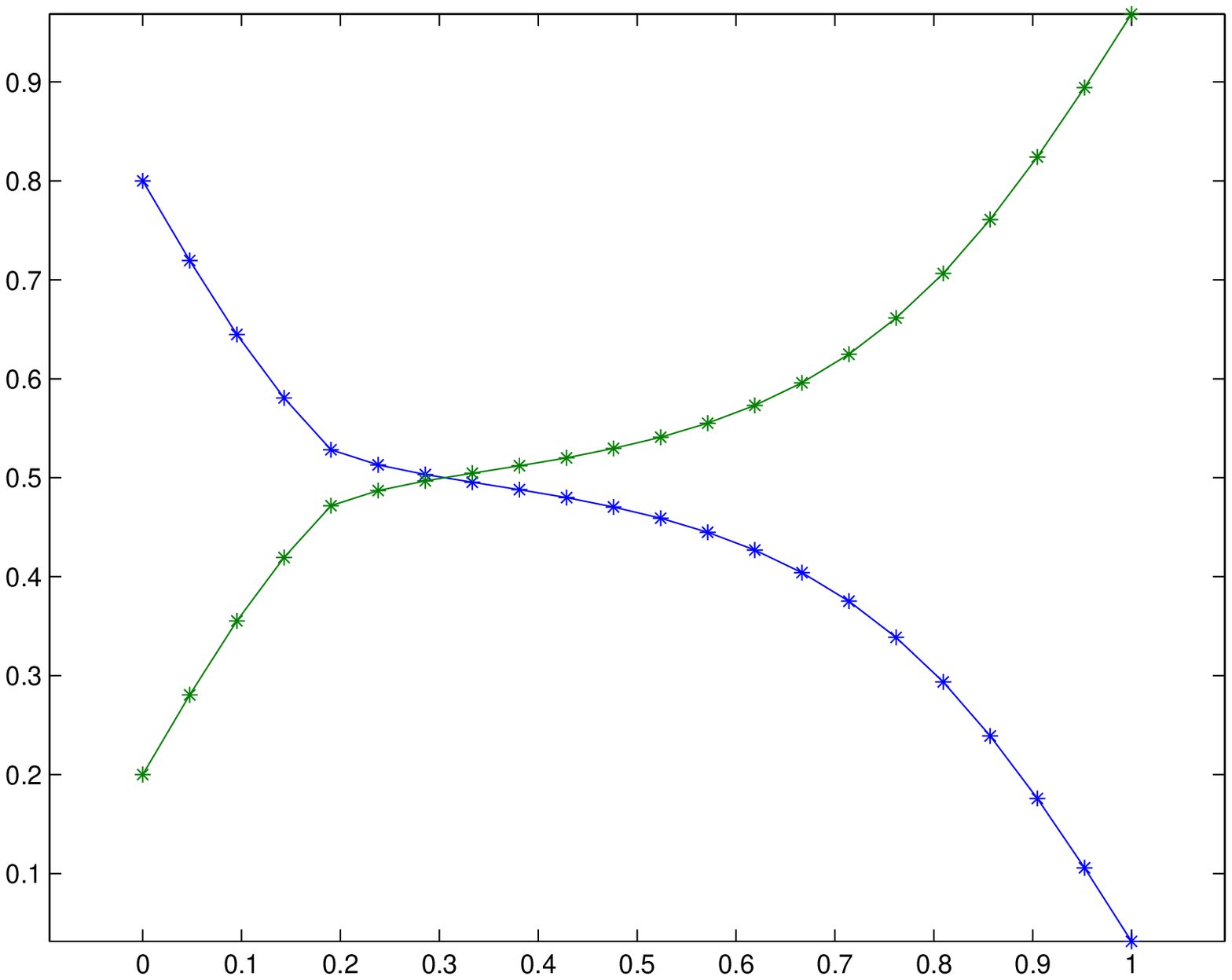}\\
\includegraphics[width=3cm]{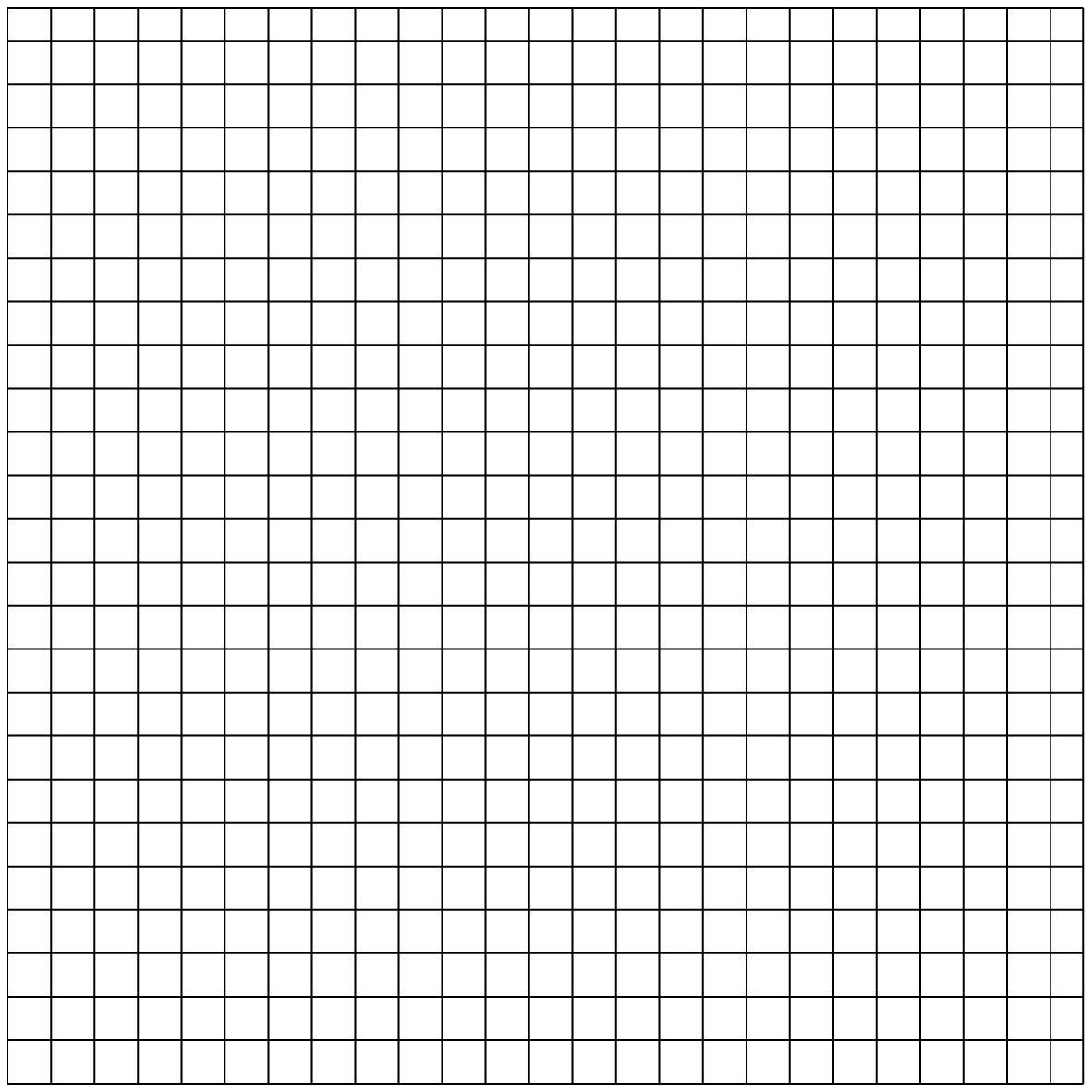}
\includegraphics[width=3cm]{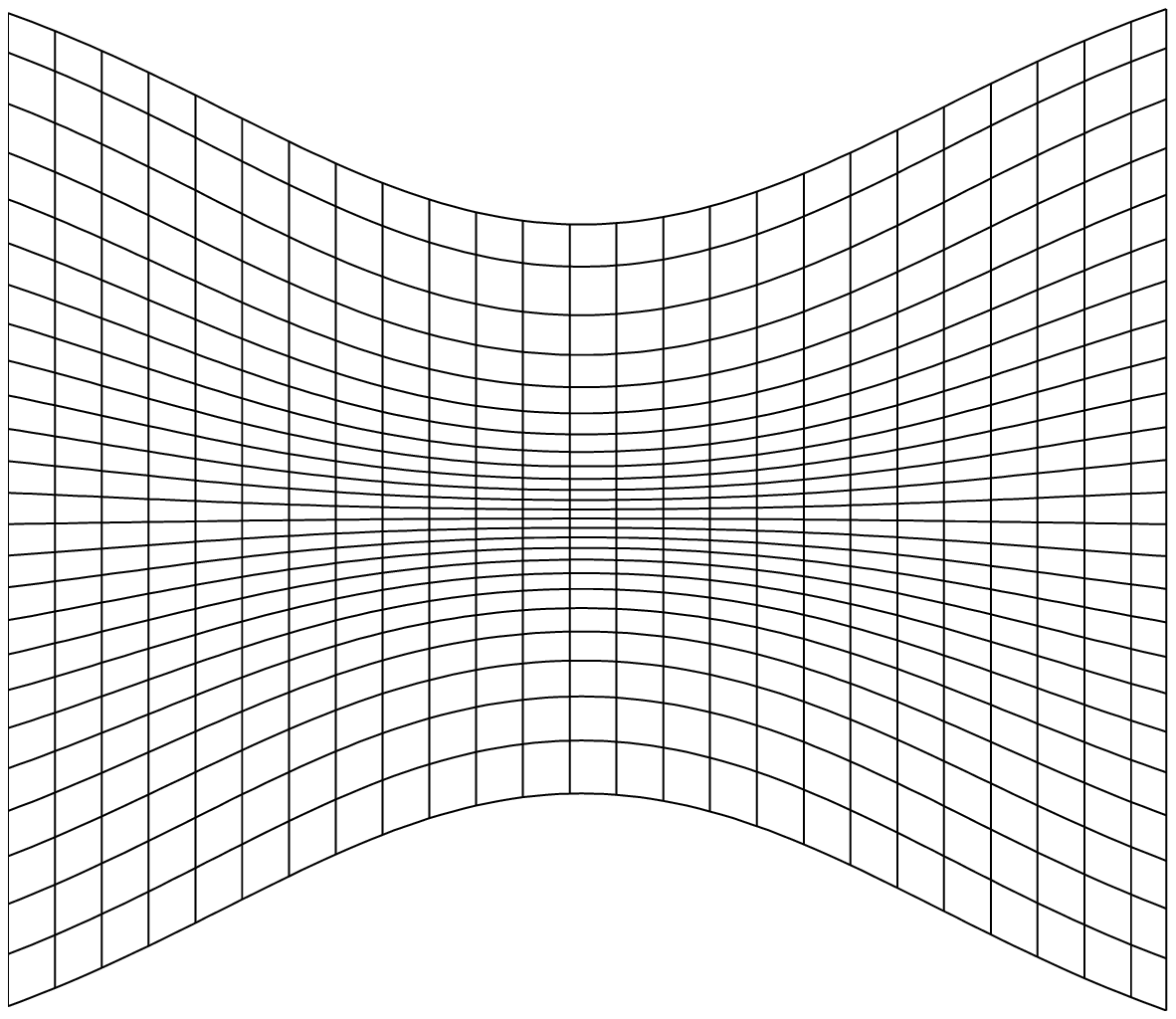}
\includegraphics[width=3cm]{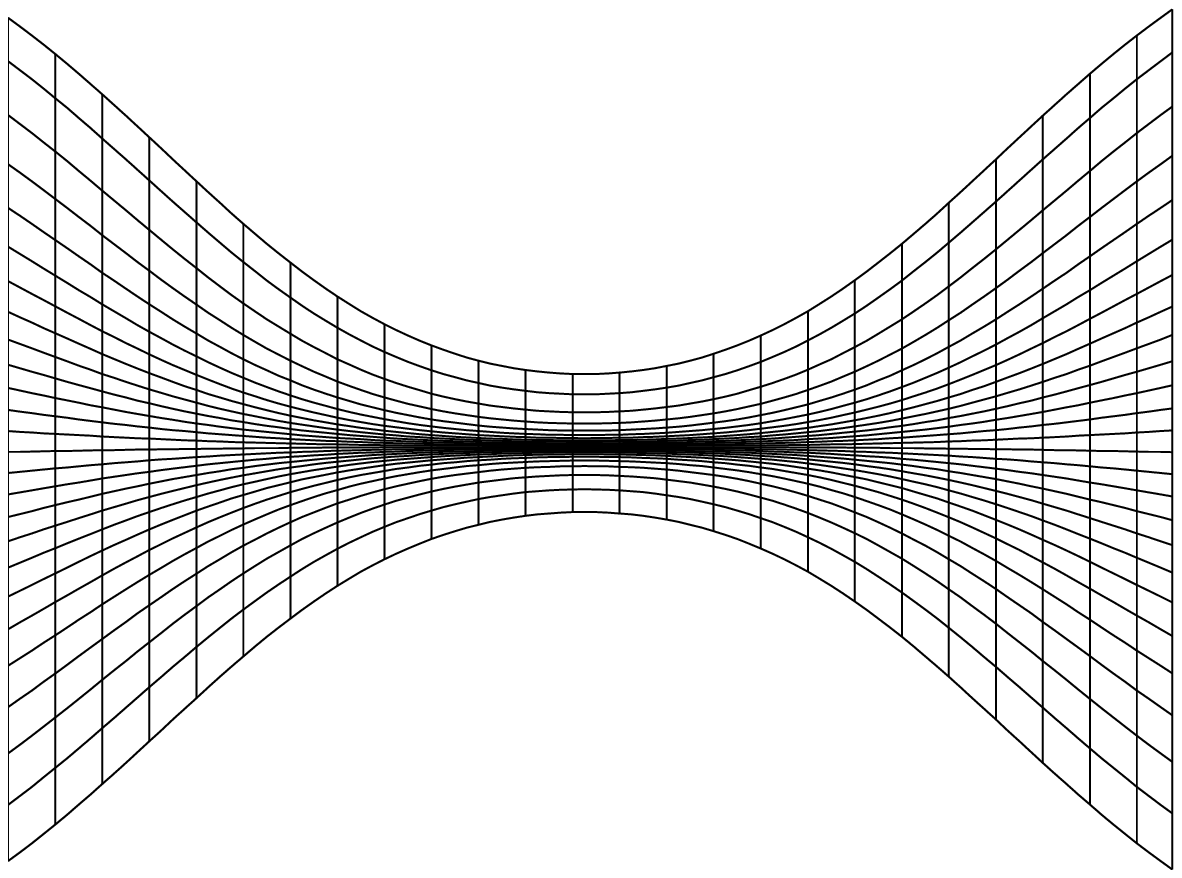}
\includegraphics[width=3cm]{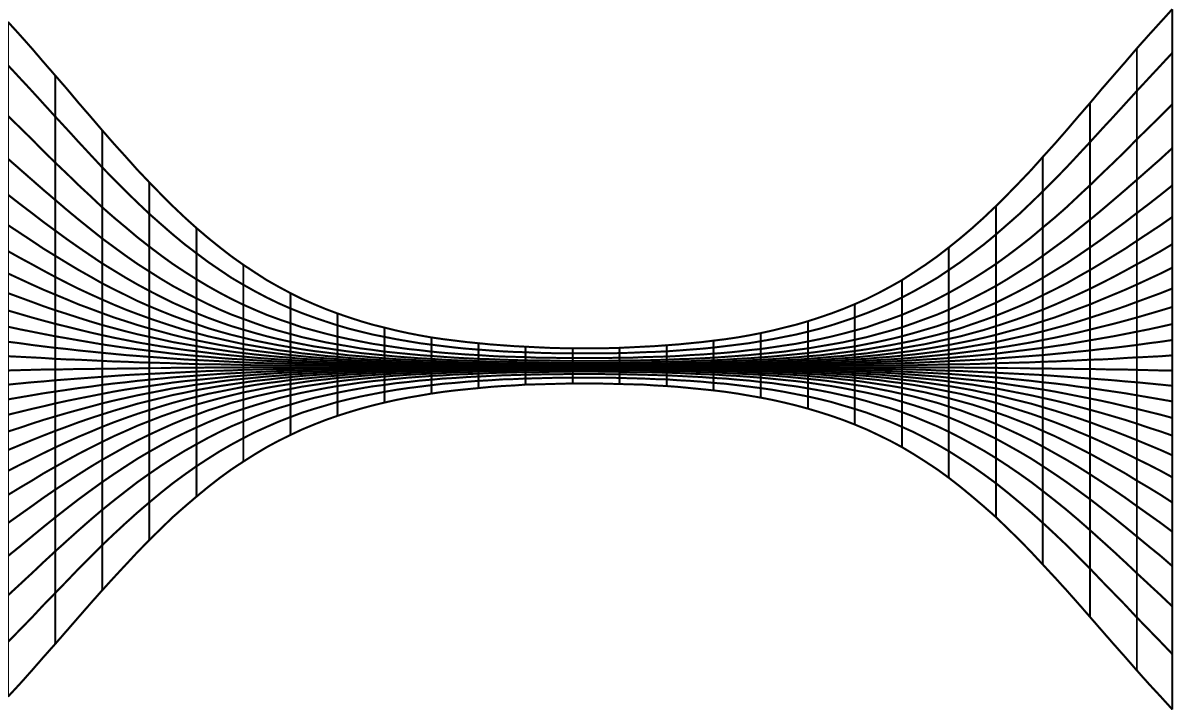}
\includegraphics[width=3cm]{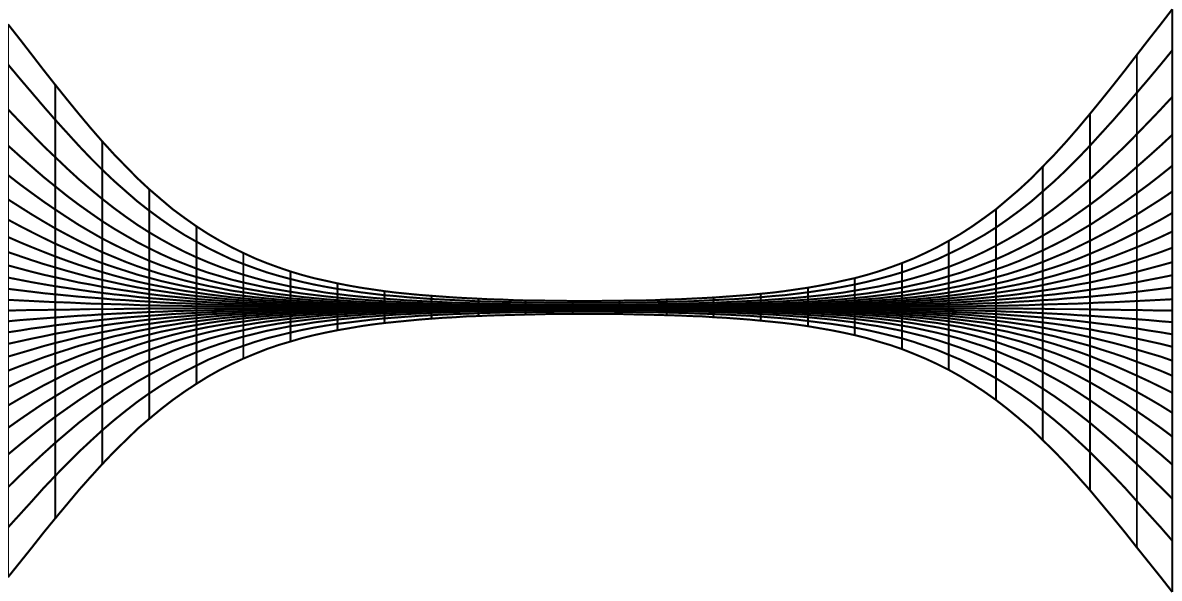}\\
\includegraphics[width=3cm]{TIFF/initPhi.eps}
\includegraphics[width=3cm]{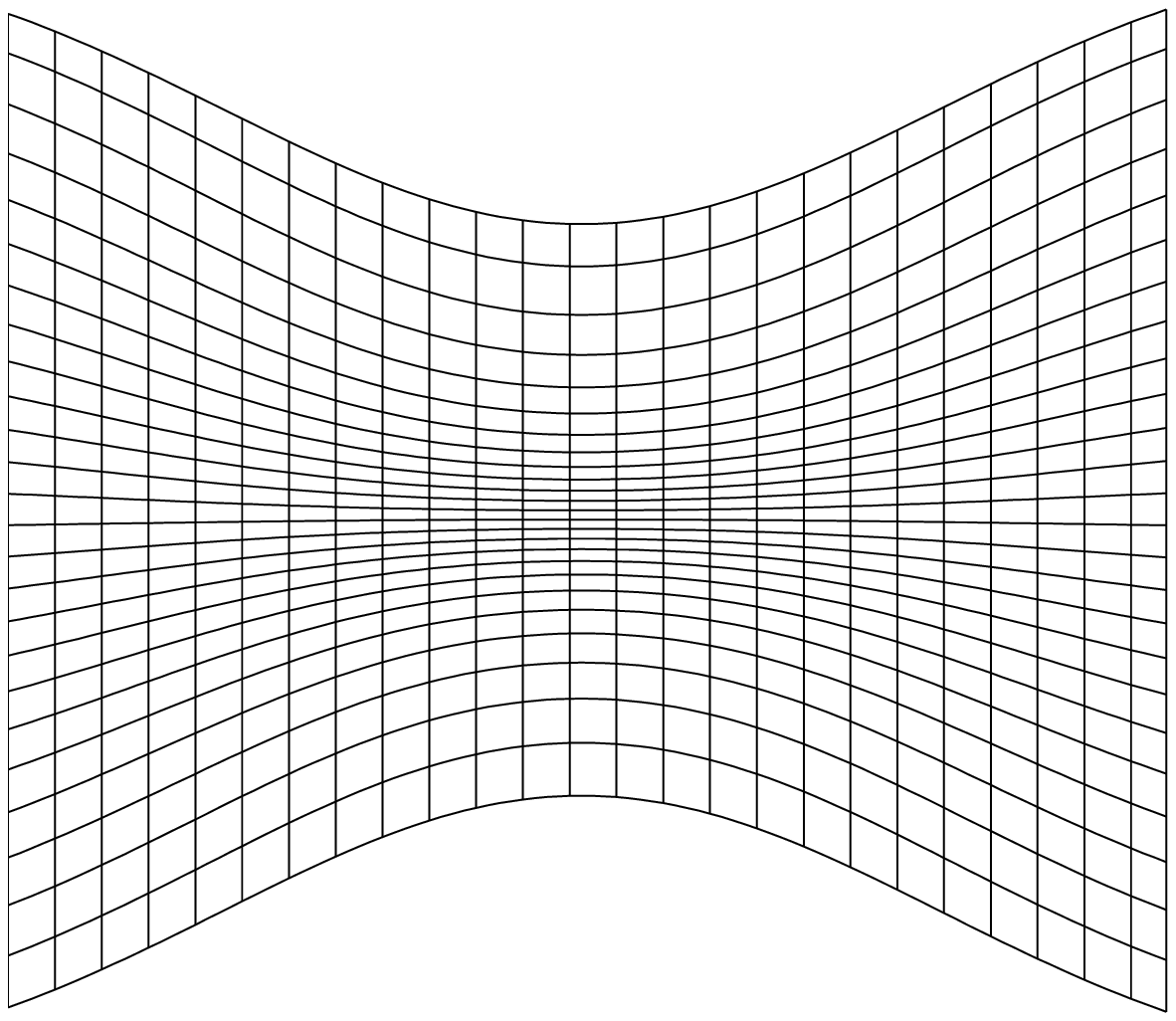}
\includegraphics[width=3cm]{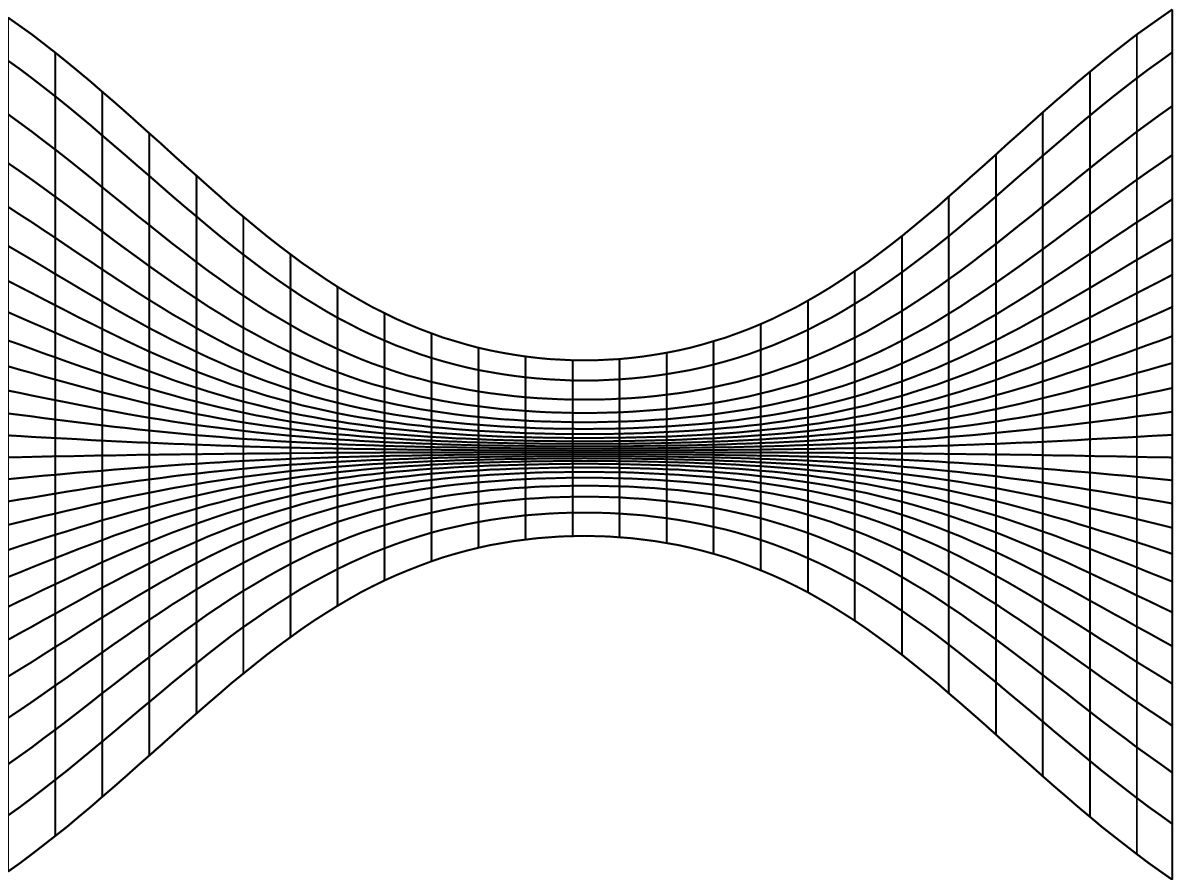}
\includegraphics[width=3cm]{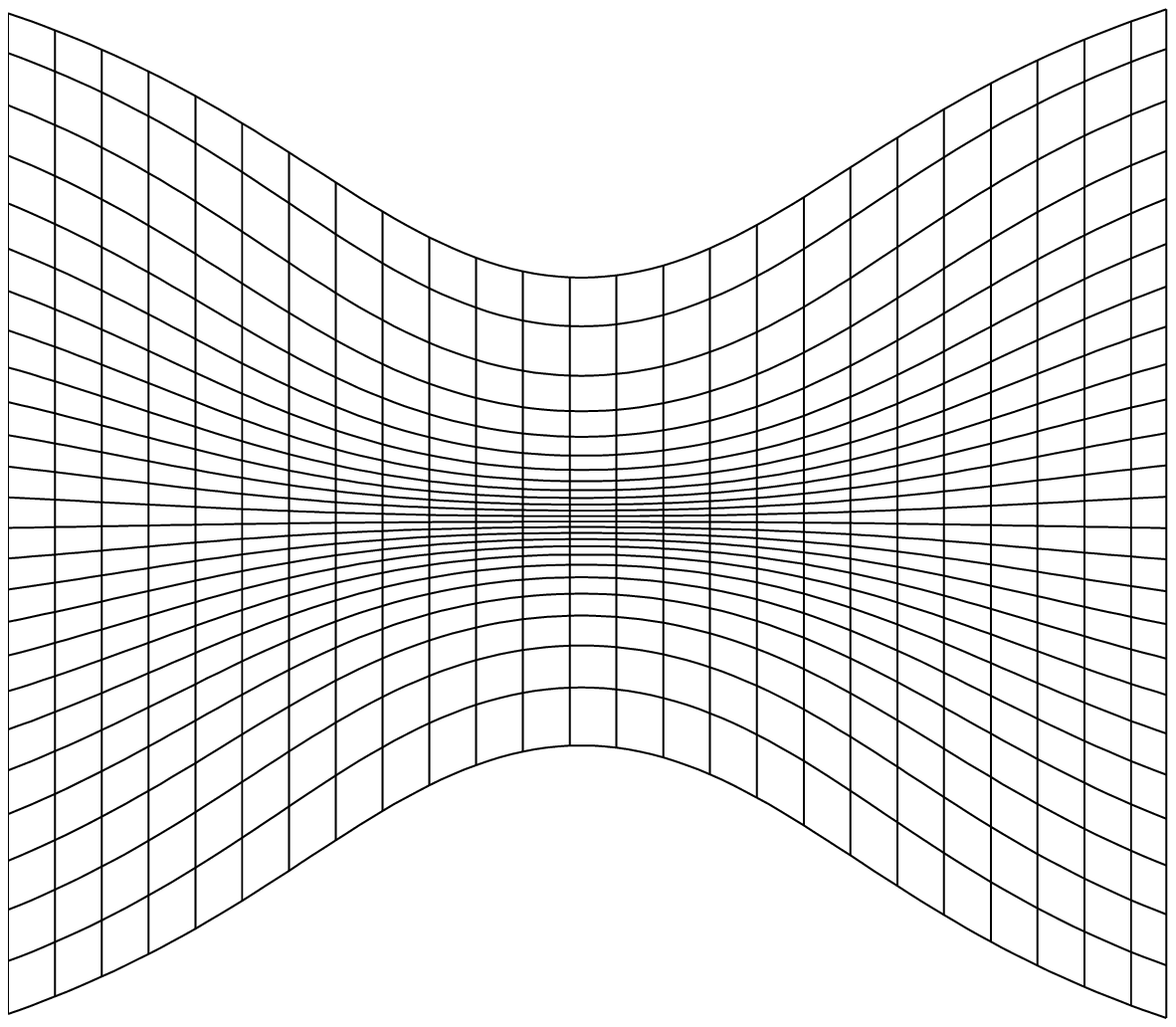}
\includegraphics[width=3cm]{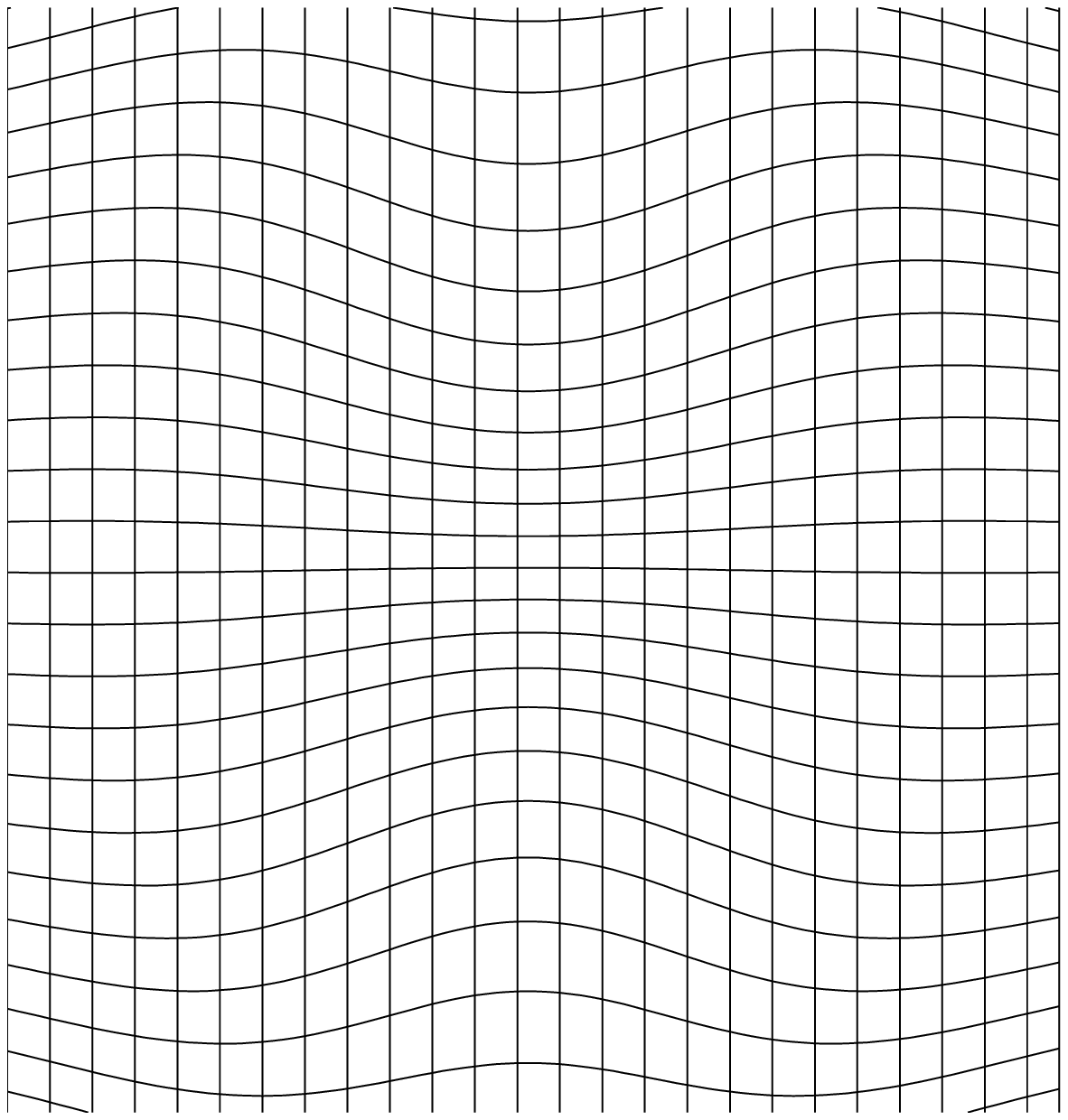}
\caption{\label{fig:1Dgauss} Deformation resulting from a head-on
  collision of two Gaussian landmarks. First row:
  $y$ coordinates of the evolving landmarks plotted against time for
  $\sigma=0$ (left) and $\sigma>0$ (right): contact requires an
infinite time when $\sigma=0$, and a crossover is observed in the
second case. In the second row, the grid deformation is plotted for
the associated 2D deformation, in the exact matching case
  ($\sigma = 0$); The grid gets squeezed while the distance between
the landmarks reduces. In this case the deformation is exactly carried
by the landmarks. The third row is with $\sigma >0$ (metamorphosis): in this
case, the landmarks travel slightly ahead of the deformation, and can cross
without creating a singularity. Before the crossover, the grid
contracts, then the landmarks diverge and the grid expands.  }
\end{center}
\end{figure*}

Interestingly enough, the crossover behavior can be recovered by using
metamorphoses, replacing $A$ by $A^\sigma$ in \eqref{eq:land.en},
because, for this model, particles are slightly disconnected from
the diffeomorphic motion, and in this precise situation slightly
ahead of it, allowing them to cross without creating a
singularity. After crossing, the landmarks carry the
diffeomorphism the other way, letting the motion appear like a
compression, then a bouncing back after the crossover.
To illustrate this, a symmetric frontal shock between two ``landmarks'' has been simulated
in the cases $\sigma=0$ vs. $\sigma > 0$ the result is in figure
\ref{fig:1Dgauss}.

\begin{figure*}
\begin{center}
\includegraphics[width=4cm]{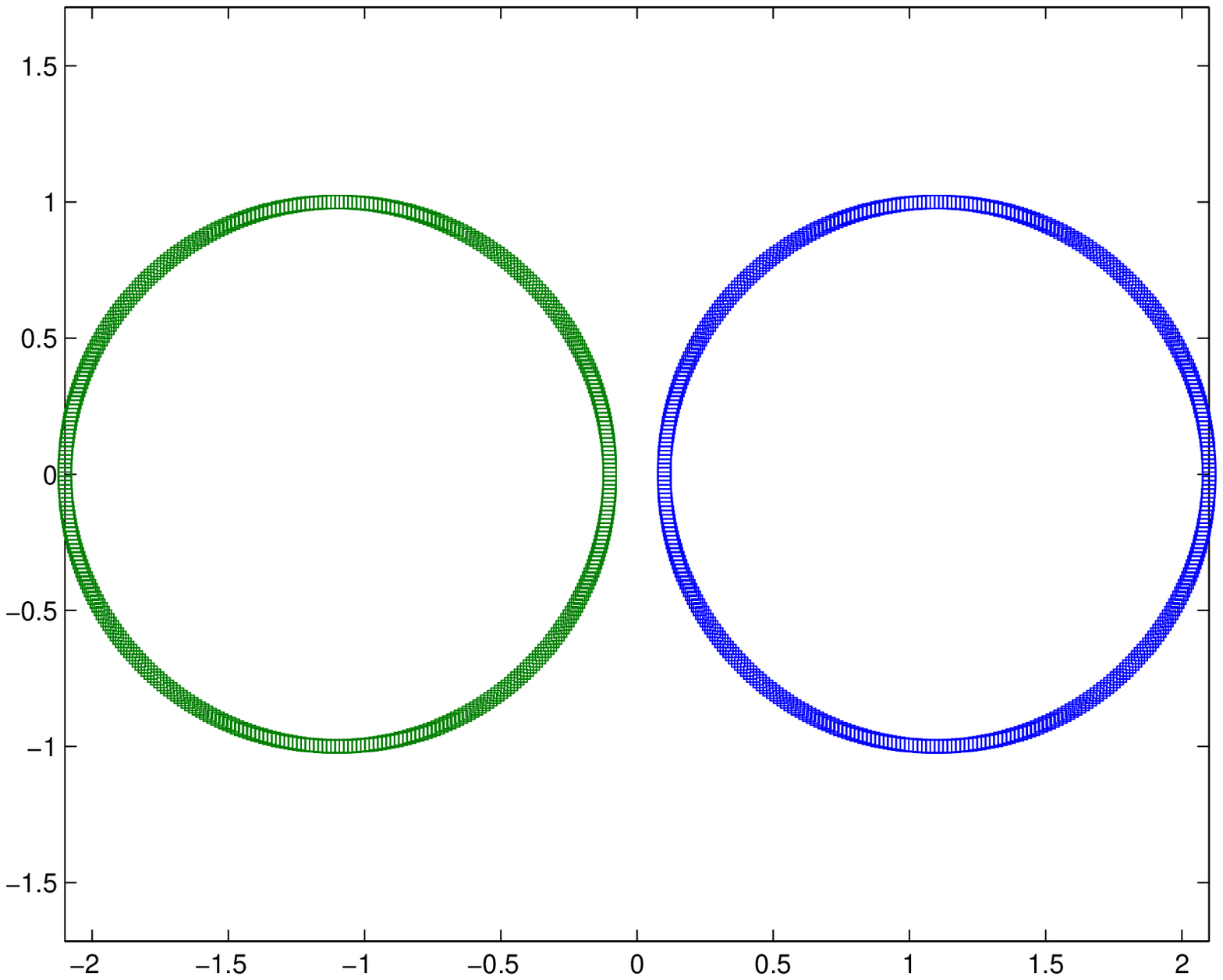}
\includegraphics[width=4cm]{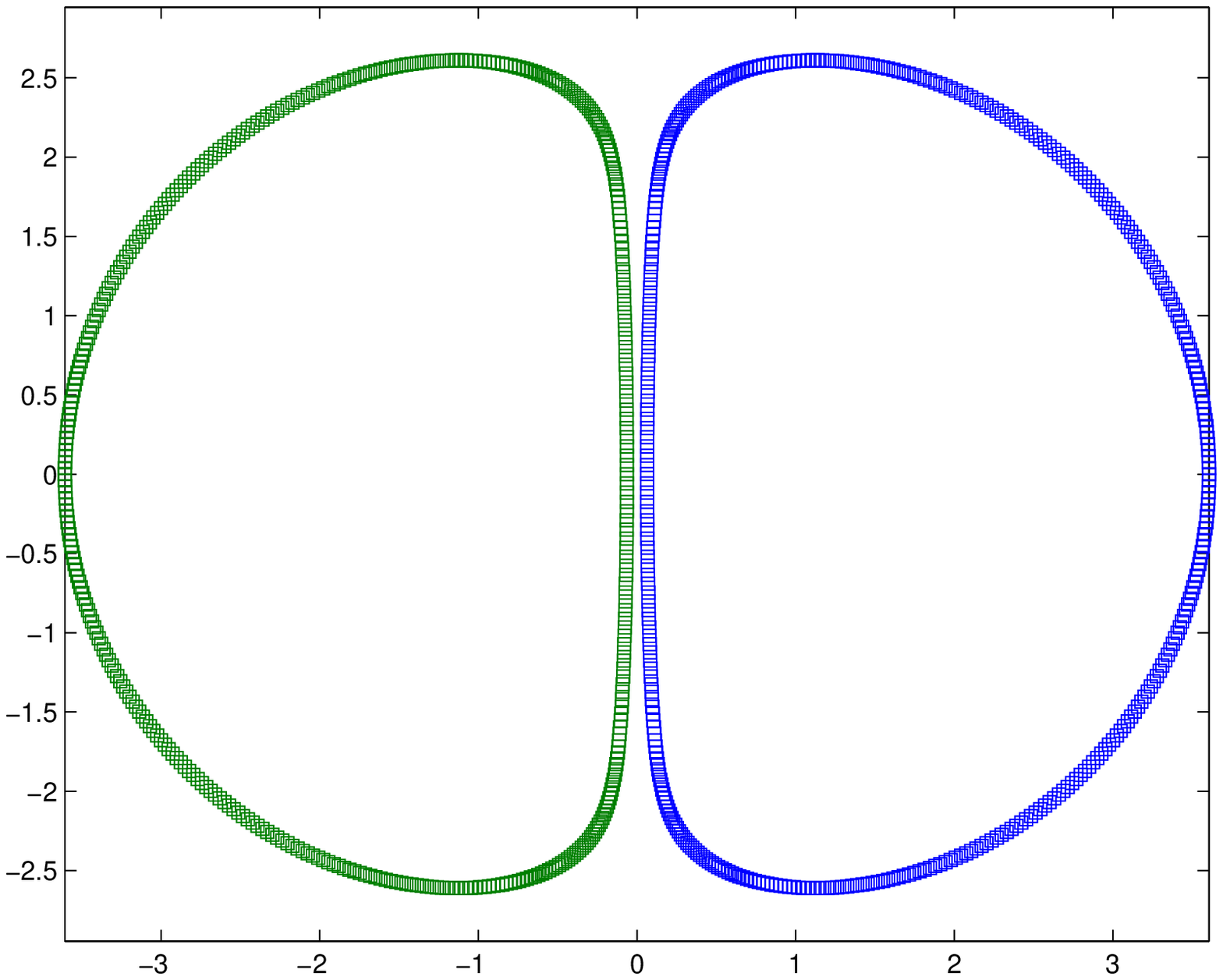}
\includegraphics[width=4cm]{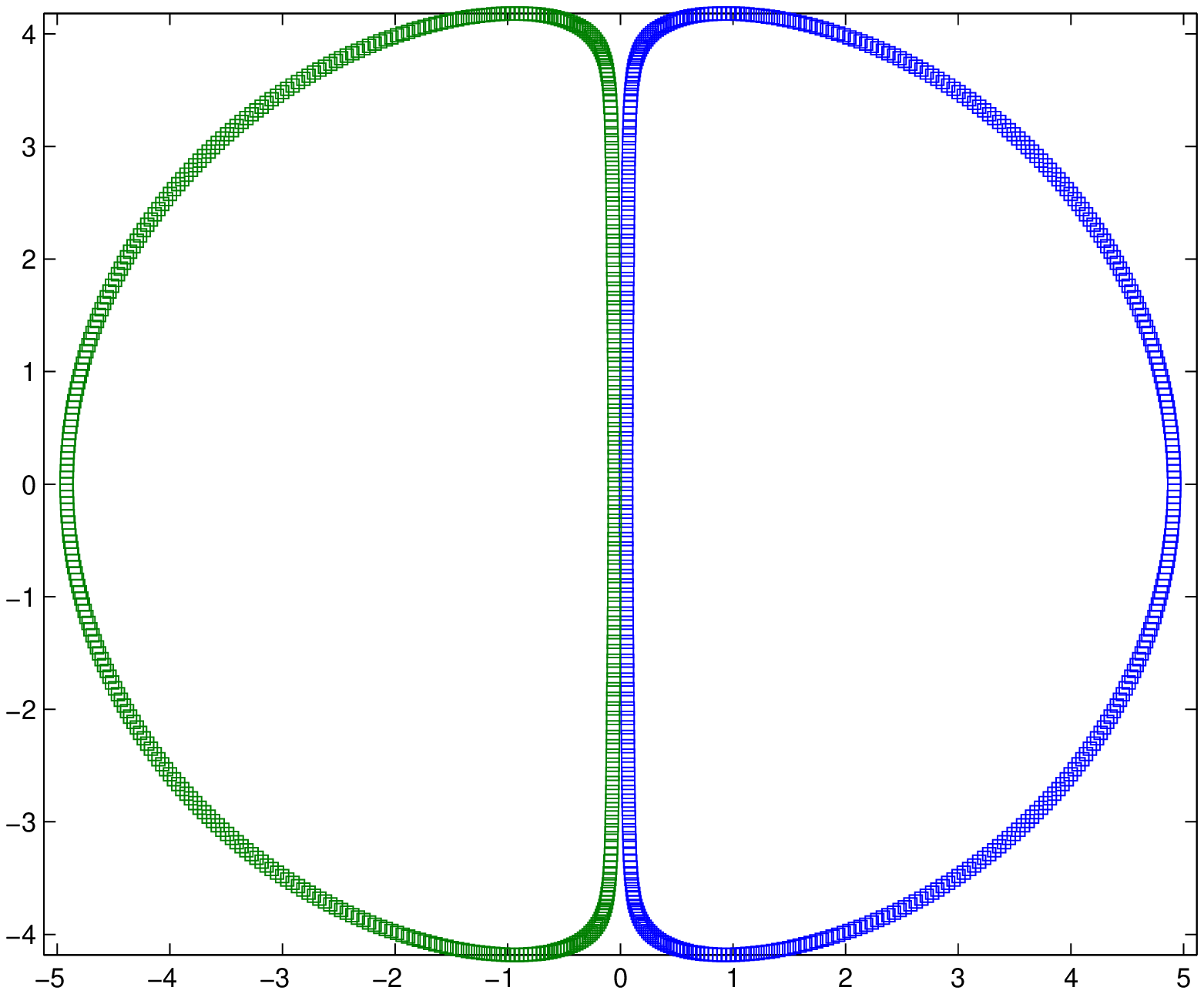}
\includegraphics[width=4cm]{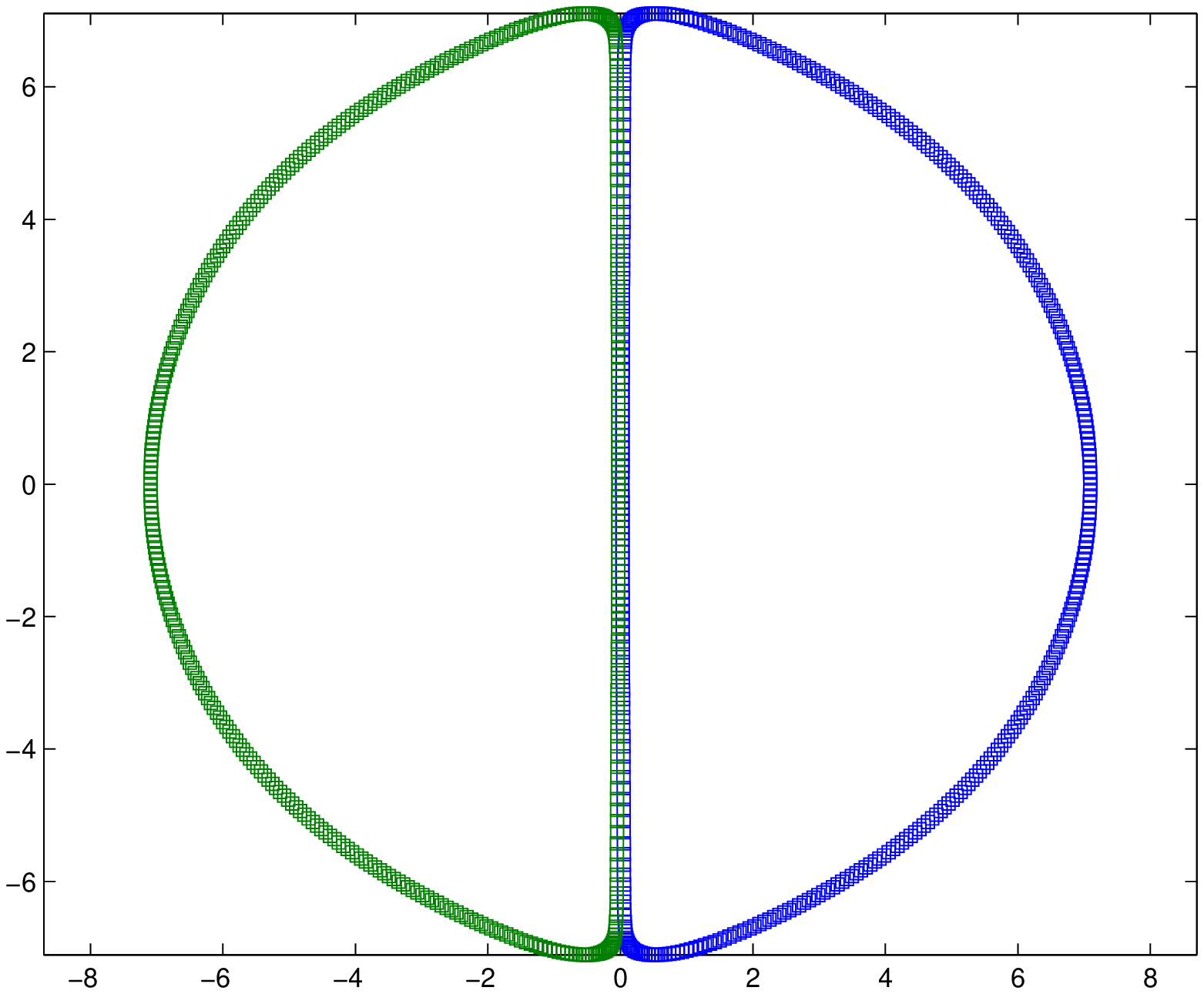}
\caption{\label{fig:twocircles0}Motion of two expanding circles
  through EPDiff with a smooth kernel. This corresponds to the case
$\sigma = 0$ (exact matching). The circles become infinitely
close, without crossing.}
\end{center}
\end{figure*}

\begin{figure*}
\begin{center}
\includegraphics[width=4cm]{TIFF/initCircles.eps}
\includegraphics[width=4cm]{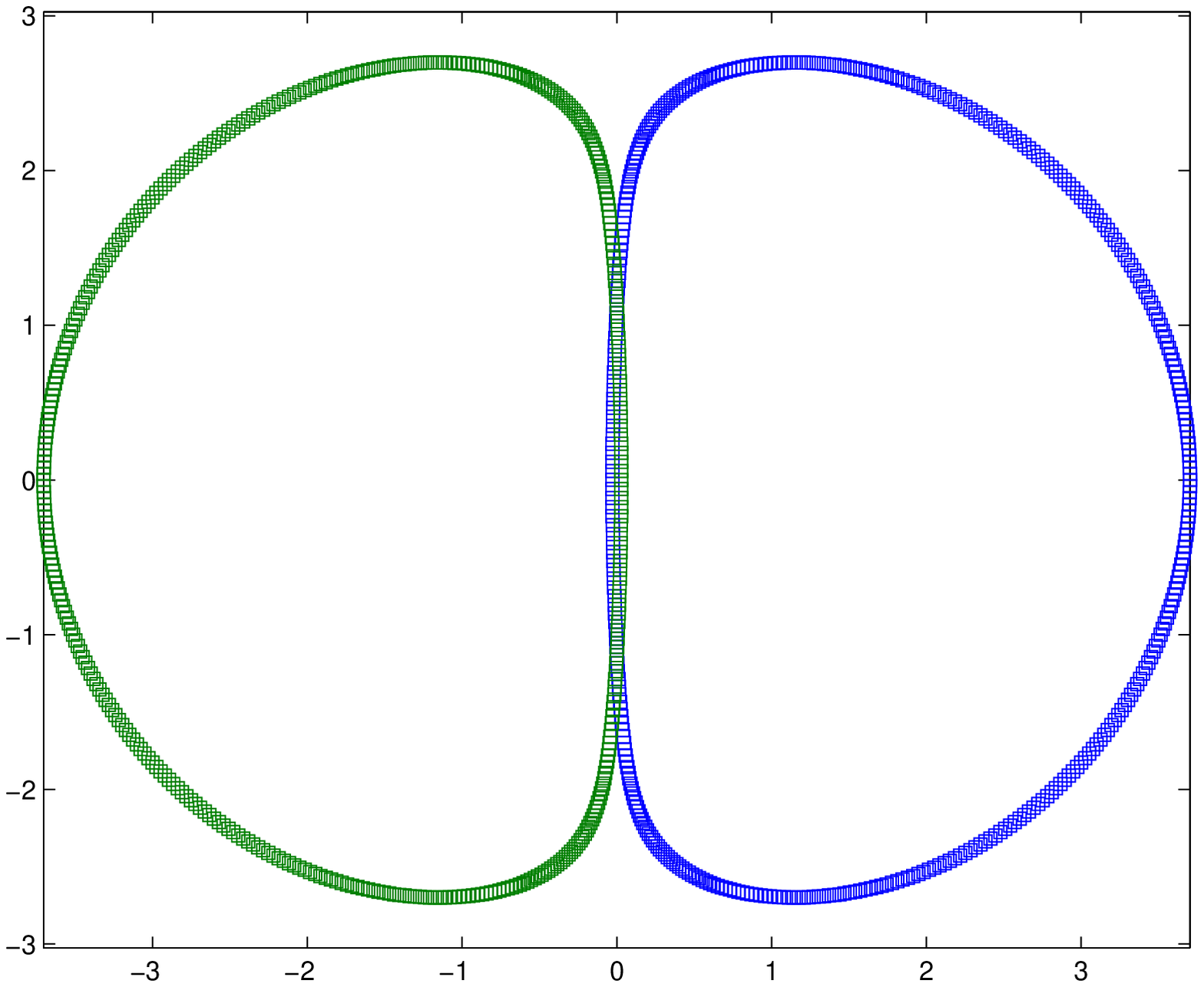}
\includegraphics[width=4cm]{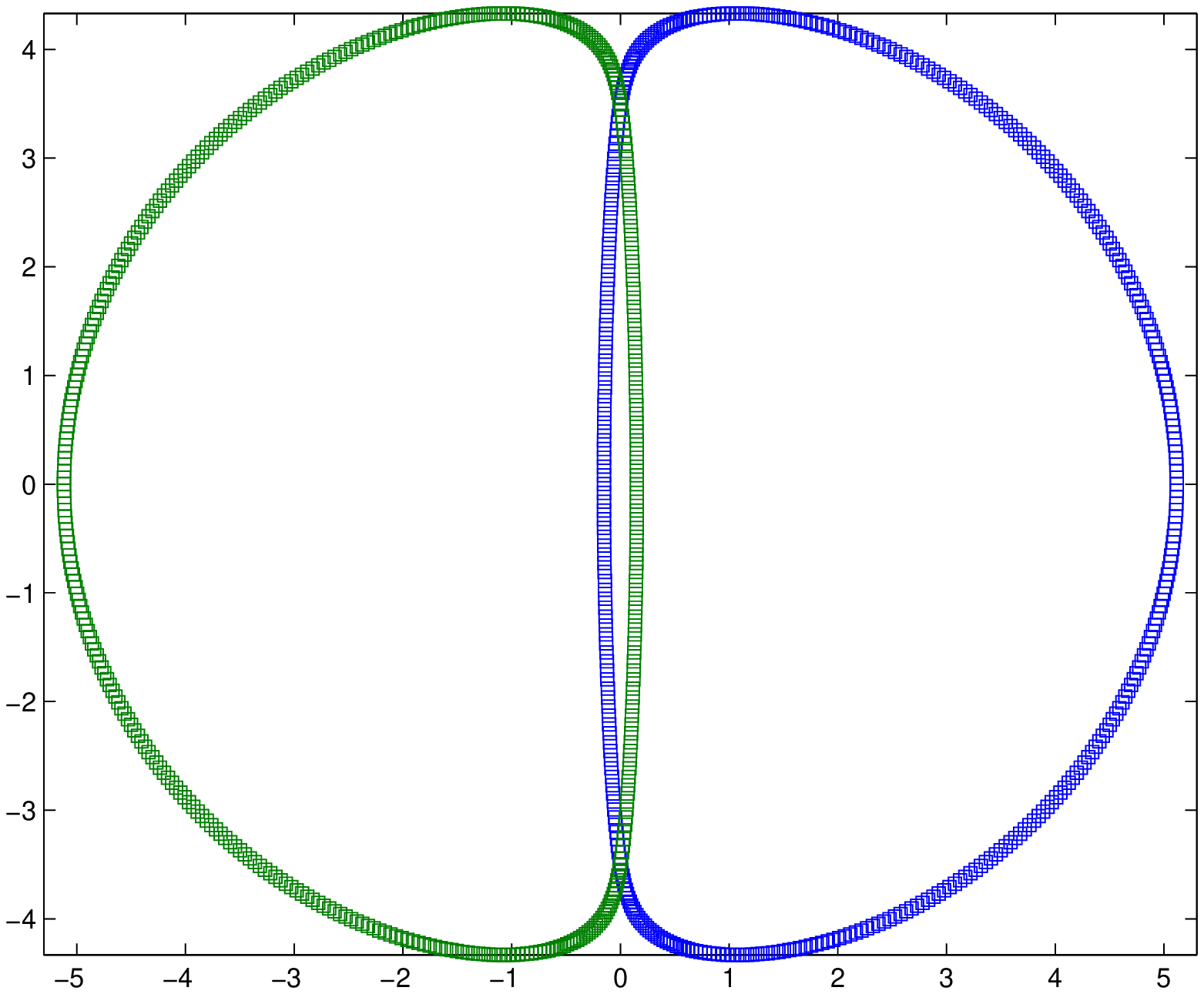}
\includegraphics[width=4cm]{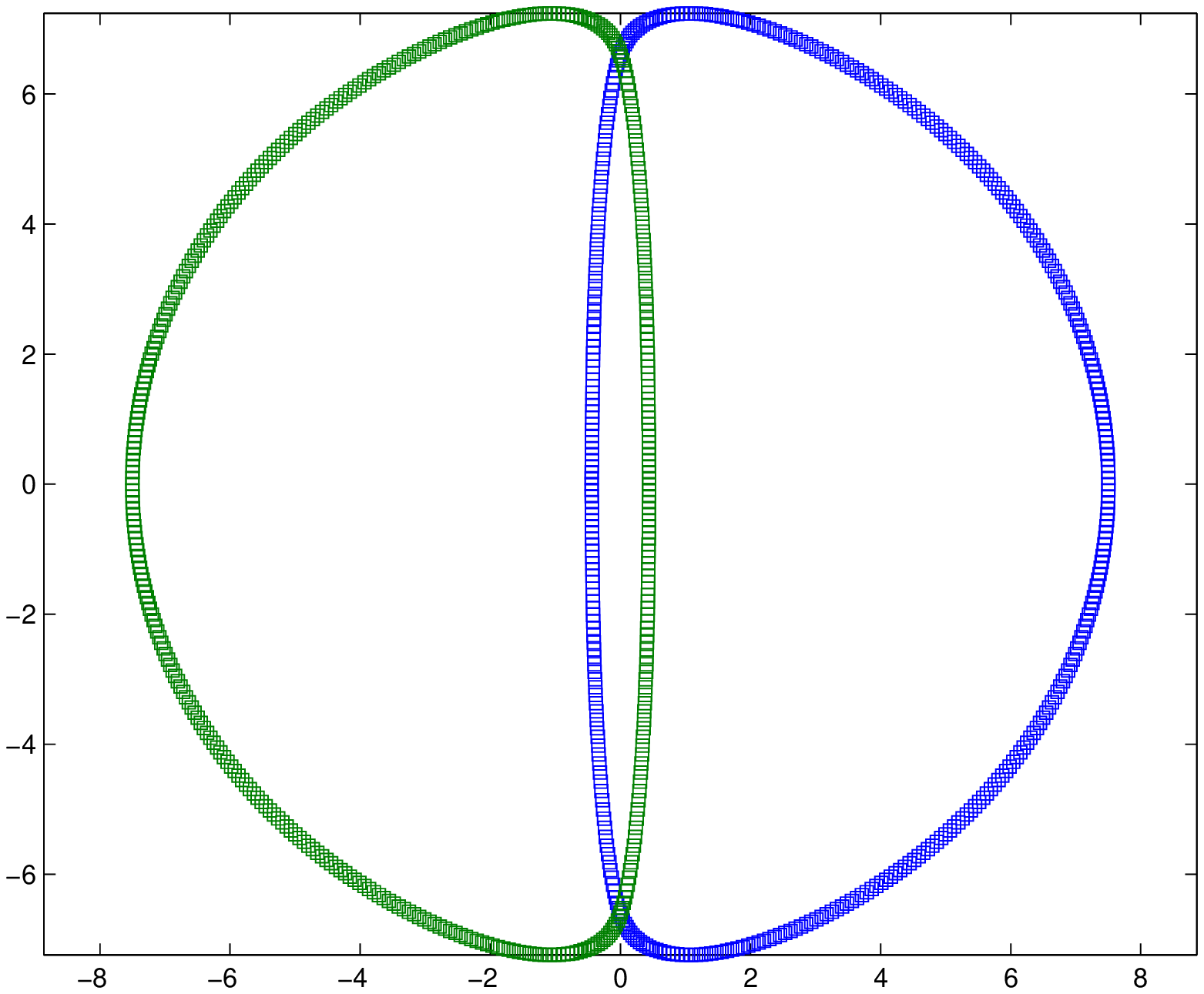}
\caption{\label{fig:twocircles1}Motion of two expanding circles
  through metamorphoses with small $\sigma^2 > 0$. The circles
intersect in final time, like the landmarks in the head-on collision
of figure \ref{fig:1Dgauss}}
\end{center}
\end{figure*}

\begin{figure*}
\begin{center}
\includegraphics[width=4cm]{TIFF/initCircles.eps}
\includegraphics[width=4cm]{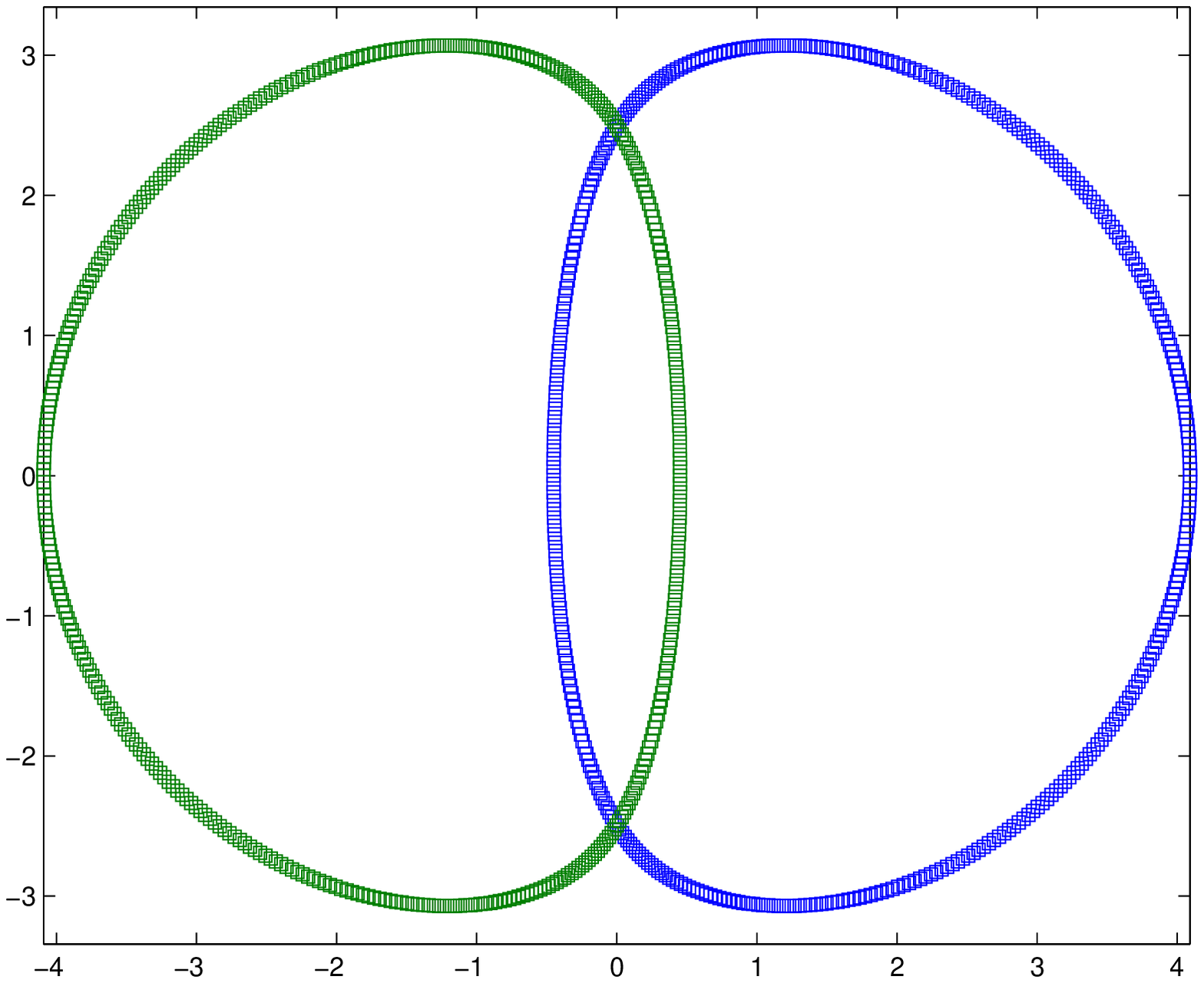}
\includegraphics[width=4cm]{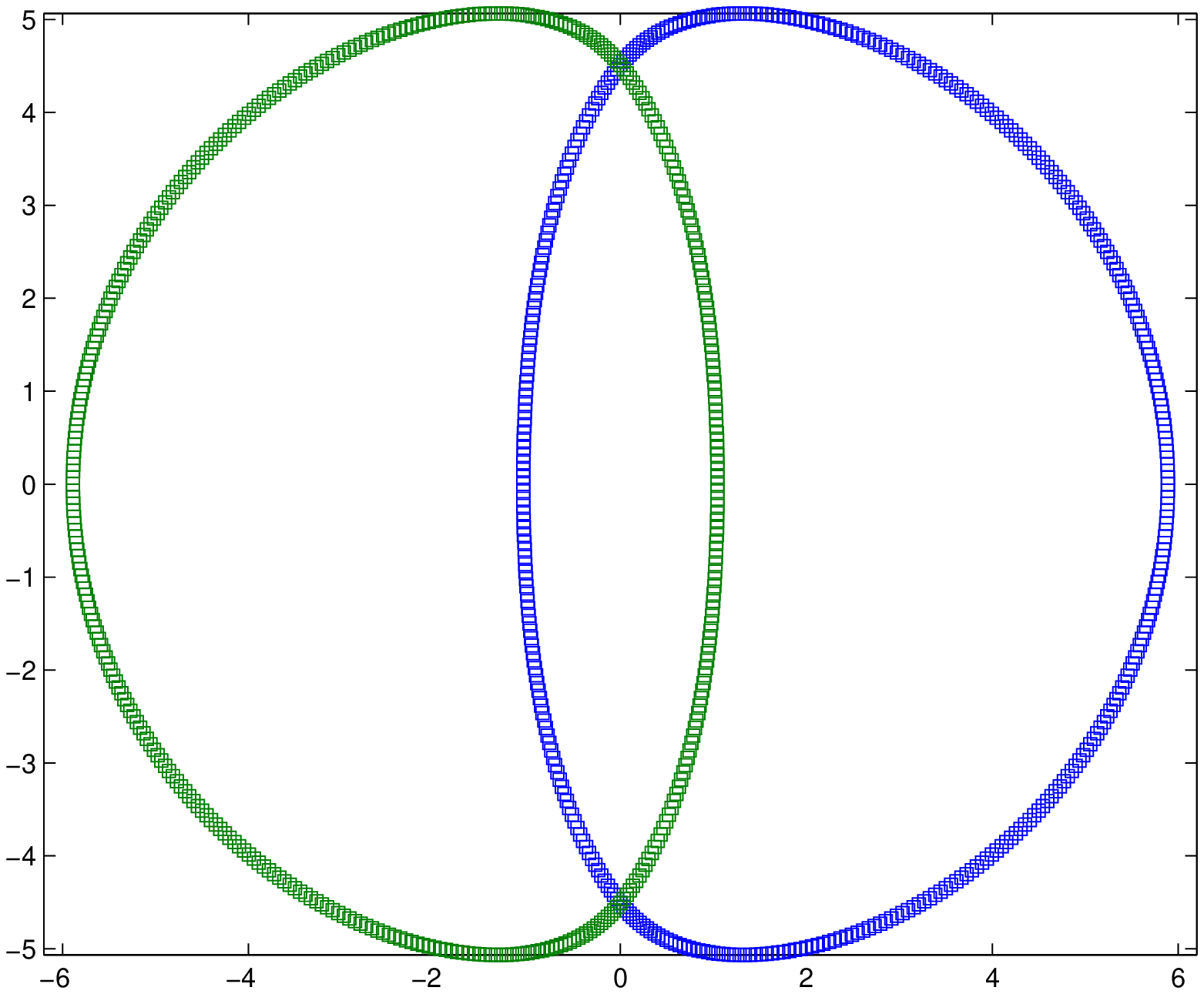}
\includegraphics[width=4cm]{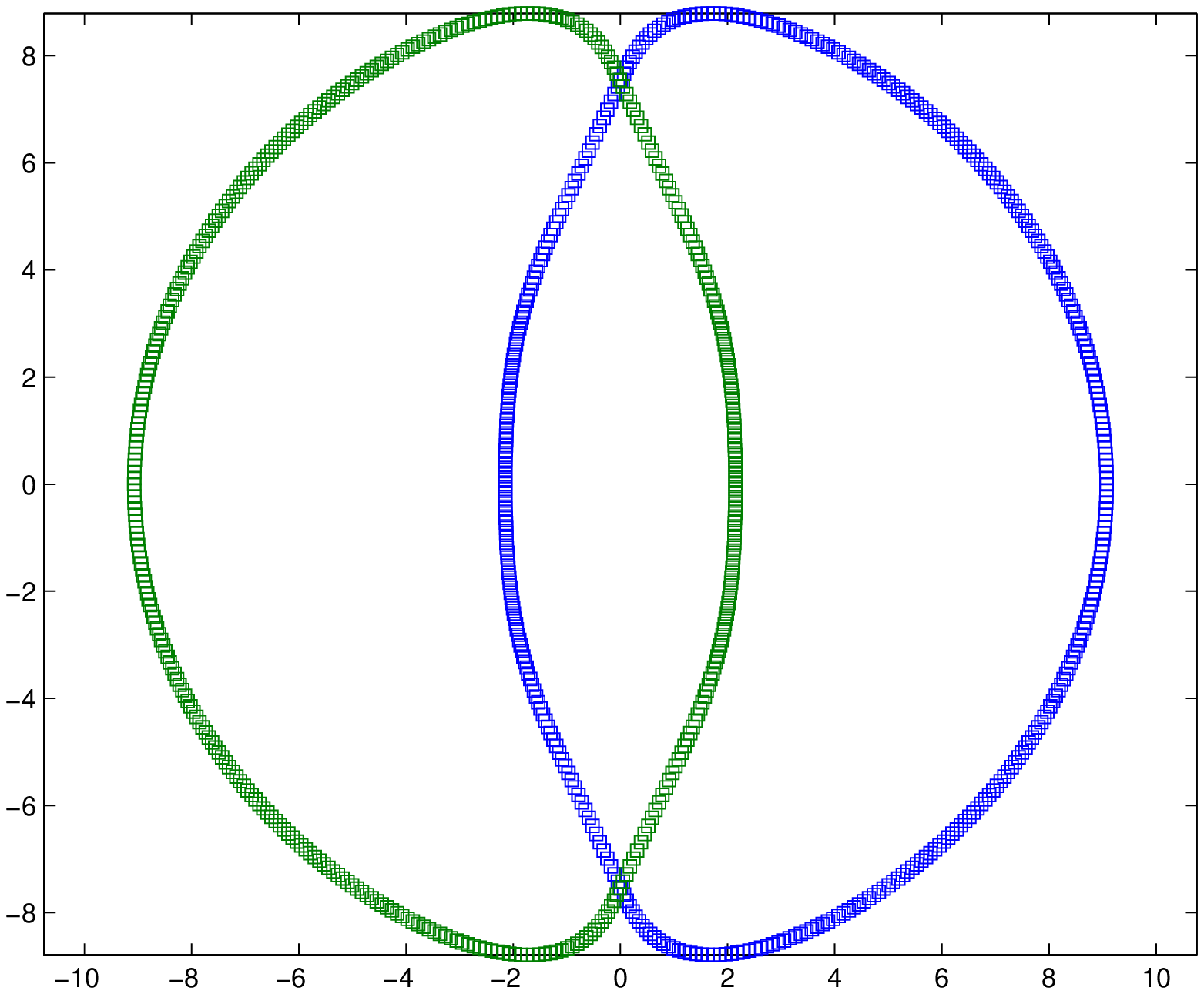}
\caption{\label{fig:twocircles2} Same as in figure
\ref{fig:twocircles1}, with a larger  $\sigma^2>0$.}
\end{center}
\end{figure*}

In 2D, the same behavior can be observed. With a smooth
kernel,
two expanding
circles which have no intersection at time $t=0$ will have no
intersection at all times. This is shown in figure \ref{fig:twocircles0}.
When the parameter
$\sigma^2$ is introduced, the circles intersect in finite time, the value
of
$\sigma^2$ influencing their shapes before and after the collision (figure \ref{fig:twocircles1} and
\ref{fig:twocircles2}).

\section{Applications of EPDiff in CA.}
\label{sec:cons}
So far,
 CA  has been related to the issue
 of comparing two geometric objects, and thus more concerned with the
 variational ``boundary value'' problem \eqref{eq:ener}. However, as
 studied in \cite{MiTrYo03}, the
 initial value problem, associated to the integration of EPDiff, turns
 out to have very important consequences for
 applications.

The main consequence of the Euler-Poincaré analysis is that, when
matching two geometric structures, the momentum at time $t=0$ contains all
the required information for reconstructing the target from the template.
This momentum therefore provides a template-centered coordinate system,
which essentially encodes all possible deformations which can be applied
to it.

There is another important feature which has been observed  in the
landmark matching problem. Although the momentum is {\it a priori}
of functional nature, as a result of the application of the
operator $L$ to the velocity field ${\mathbf{u}}_t$, in the landmark
case, we found that it was characterized by a collection of $K$
vectors in space, for a matching problem with $K$ landmarks. Thus,
the momentum has exactly the same dimension as the matched
structures, and there is no redundancy of the representation. This
is generic in the sense that the final dimension of the momentum
exactly adapts to the nature of the matching problem. For example,
as stated in \cite{MiTrYo03, MiTrYo03b}, when matching two smooth scalar
images, the momentum has to be normal to the level sets of the
image, and therefore is uniquely characterized by its intensity.
The momentum can therefore be seen as an algebraic image, which
has the same dimension as the original image. This is because
template matching brings an additional reduction to the original
analysis of finding optimal paths between diffeomorphisms (which
leads to the EPDiff equation). This reduction is because the solution must
be modded out by the diffeomorphisms that leave the template invariant,
which constrains the optimal solution. As a  result, the
momentum is a (locally) one-to-one representation of the targets, in this
template-based coordinate system. It is therefore a non-redundant tool for
representing deformations of the template.

Besides being one-to-one, the other advantage of the momentum
representation is that it is linear in nature, being dual to the velocity
vectors. Thus, linear combinations of either velocity fields, or momenta
are meaningful mathematically and physically, provided they are applied to
the same template. Thus the average of a collection of
momenta, of their principal components, or time derivatives of momenta at
a fixed template are all well-defined quantities.

\begin{figure*}
\begin{center}
\includegraphics[width=5cm]{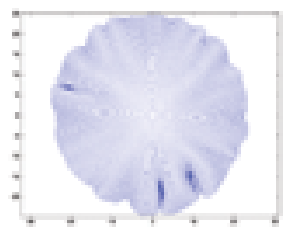}
\includegraphics[width=5cm]{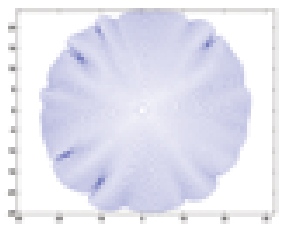}
\includegraphics[width=5cm]{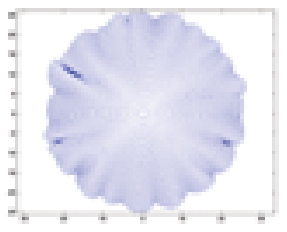}
\caption{\label{fig:random} Three deformations of a disc by EPDiff under
random initial conditions. These figures visualize the evolution of a disc
under the evolution of EPDiff, for initial uncorrelated noise momentum on
its boundary. This shows how random deformations in momenta superpose
linearly to produce a diffeomorphic change in shape.}
\end{center}
\end{figure*}

\begin{figure*}
\begin{center}
\includegraphics[width=6.4cm]{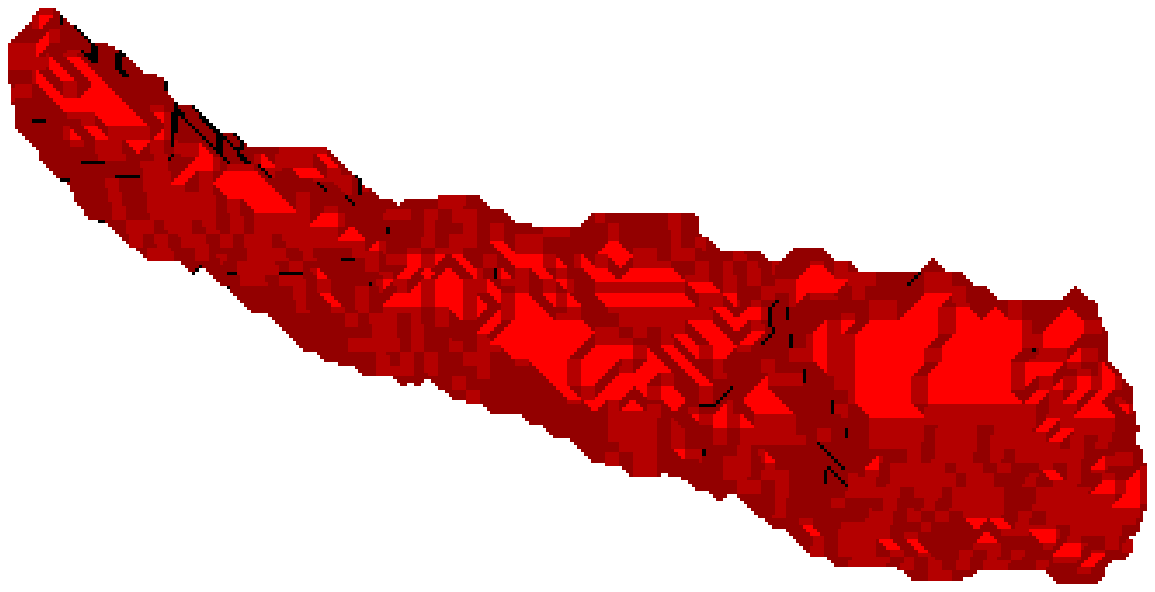}
\includegraphics[width=6.4cm]{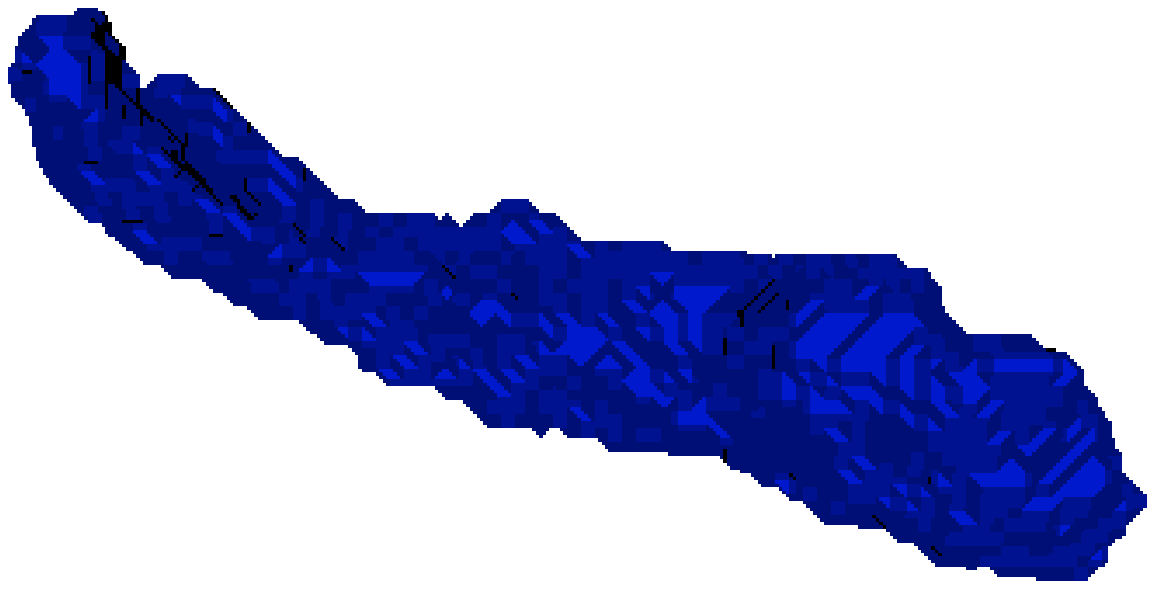}\\
\includegraphics[width=6.4cm]{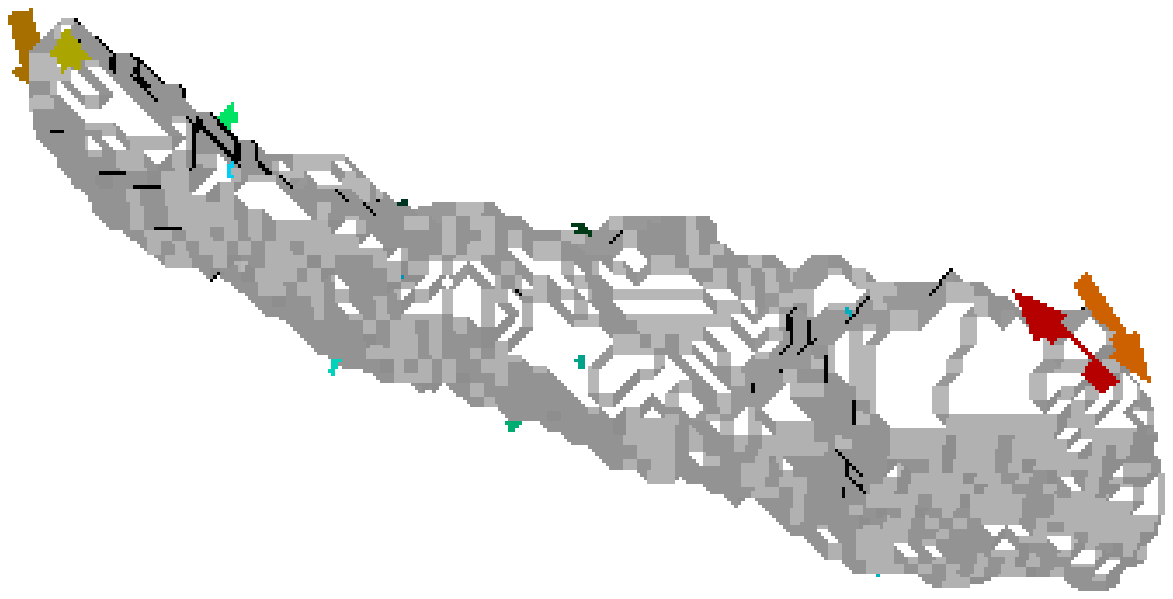}
\includegraphics[width=6.4cm]{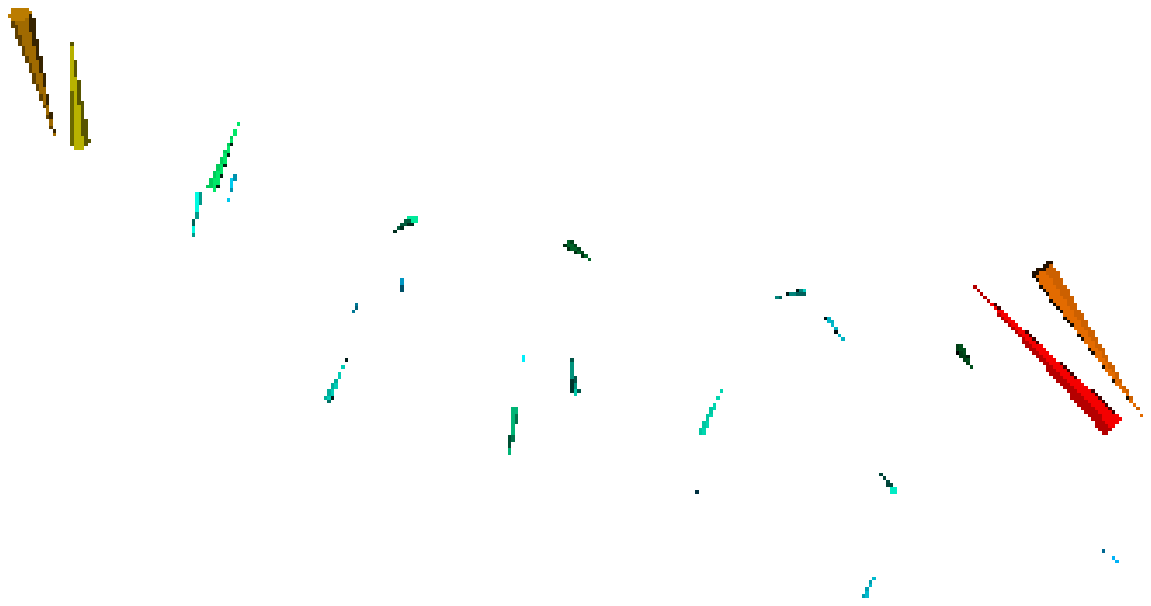}
\caption{\label{fig:hippo} Landmark matching for hippocampi. Two
sets of landmarks have been manually defined on the hippocampi in
the first row, the first one being considered as the template. The
second row shows the initial momentum superimposed with the
template (left) and alone (right). Most of the momenta is
concentrated at the head and tail of the hippocampus. Data taken
from the Biomedical Informatics Research Network
(\protect\url{www.nbirn.net})}
\end{center}
\end{figure*}

Finally, because they provide an efficient tool for representing
deformable data, momenta are perfectly suitable for modeling
deformations. Any statistical model on momenta provides, after the
integration of EPDiff, a statistical model on deformations, the
advantage being that it is much easier to build, sample and
estimate statistical models on a linear space. An illustration of
this is provided in figure \ref{fig:random}, in which the momentum
was generated as positive uncorrelated noise on the boundary of a
disc, and EPDiff was integrated with this initial conditions. The
Green's function we used for this is a Gaussian kernel, $G(x,y) =
\exp(-|x-y|^2)$. Figure \ref{fig:hippo} shows the initial momentum of
the geodesic path between two 3D sets of landmarks placed on two
hippocampi. We also refer to the statistical experiments on
PCA in momentum space presented in \cite{mtvy04}.

\section{Conclusions}\label{conc-sec}

We have identified momentum as a key concept in the representation
of image data for CA, and discovered important analogies with
soliton dynamics. Future work will explore further applications of
EPDiff. In particular, it will be interesting to see whether the
{\it exchange} of momentum in the interactions of multiple
outlines will become as useful a concept in computational image
analysis as it is in soliton interaction dynamics.

{\bf Acknowledgments.}
DDH is grateful for partial support by US DOE, under contract
W-7405-ENG-36 for Los Alamos National Laboratory, and Office of
Science ASCAR/AMS/MICS. JTR is grateful for support from NIH
(MH60883,MH 62130,P41-RR15241-01A1, MH62626, MH621130-01A1, MH064838,
P01-AG03991-16), NSF NPACI and NOHR.

\bibliographystyle{abbrv}
\bibliography{solitonsInCA}

\end{document}